\documentclass[12pt]{iopart}

\usepackage{iopams}
\usepackage{graphicx}
  \expandafter\let\csname equation*\endcsname\relax
  \expandafter\let\csname endequation*\endcsname\relax
\usepackage[12pt]{extsizes}
\usepackage[T1]{fontenc}
\usepackage[francais,english]{babel}
\usepackage{layout}
\usepackage[margin=1in]{geometry}
\usepackage{multicol}
\usepackage{float}
\restylefloat{table}
\usepackage{amsthm}
\usepackage{amsmath}
\usepackage{amssymb}
\usepackage{mathrsfs}
\usepackage{stmaryrd} 
\numberwithin{equation}{section}
\usepackage[usenames]{color}

\usepackage[backref=false]{hyperref}
\hypersetup{
colorlinks=true,
citecolor=red,
linkcolor=darkblue
}
\definecolor{darkblue}{rgb}{0,0,.8}
\definecolor{red}{rgb}{1,0,0}

\usepackage[vcentermath]{youngtab}
\usepackage{ytableau}

\newcommand{\nc}{\newcommand}
\nc{\be}{\begin{equation}}
\nc{\ee}{\end{equation}}
\nc{\bea}{\begin{eqnarray}}
\nc{\eea}{\end{eqnarray}}
\nc{\la}{\lambda}
\nc{\sg}{\sigma}
\nc{\dt}{\delta}
\nc{\Sc}{\mathcal{S}}
\nc{\Nc}{\mathcal{N}}
\nc{\rc}[1]{\textcolor{red}{#1}}
\newcommand{\ex}{\mathrm{e}}
\nc{\sn}[1]{\tilde{#1}}
\nc{\sya}[1]{\hat{#1}}
\nc{\Op}{\mathcal{O}}
\usepackage{tikz}
\usetikzlibrary{calc}
\usetikzlibrary{chains}
\usetikzlibrary{scopes}
\usetikzlibrary{shapes.geometric}
\usetikzlibrary{trees}
\usetikzlibrary{topaths}
\usetikzlibrary{positioning}
\usetikzlibrary{patterns,decorations.pathreplacing}

\newcommand{\Pnn}{{\mathbb P}\left(
\begin{tikzpicture}[baseline=8pt]
 \draw [fill] (0,0) circle(0.2ex);
 \draw [fill] (0.2,0) circle(0.2ex);
 \draw [fill] (0,0.85) circle(0.2ex);
 \draw [fill] (0.2,0.85) circle(0.2ex);
 \end{tikzpicture}
\right)}

\newcommand{\Ps}{{\mathbb P}\left(
\begin{tikzpicture}[baseline=8pt]
 \draw [fill] (0,0) circle(0.2ex);
 \draw [fill] (0.2,0) circle(0.2ex);
 \draw [fill] (0,0.85) circle(0.2ex);
 \draw [fill] (0.2,0.85) circle(0.2ex);
 \draw [thick] (0,0)--(0,0.85);
 \draw [thick] (0.2,0)--(0.2,0.85);
 \end{tikzpicture}
\right)}

\newcommand{\Pc}{{\mathbb P}\left(
\begin{tikzpicture}[baseline=8pt]
 \draw [fill] (0,0) circle(0.2ex);
 \draw [fill] (0.2,0) circle(0.2ex);
 \draw [fill] (0,0.85) circle(0.2ex);
 \draw [fill] (0.2,0.85) circle(0.2ex);
 \draw [thick] (0,0)--(0.2,0.85);
 \draw [thick] (0.2,0)--(0.,0.85);
 \end{tikzpicture}
\right)}

\newcommand{\Panan}{{\mathbb P}\left(
\begin{tikzpicture}[baseline=8pt]
 \draw [fill] (0,0) circle(0.2ex);
 \draw [fill] (0.2,0) circle(0.2ex);
 \draw [fill] (0,0.85) circle(0.2ex);
 \draw [fill] (0.2,0.85) circle(0.2ex);
 \draw [thick] (0,0)--(0,0.85);
 \end{tikzpicture}
\right)}

\newcommand{\Pnbnb}{{\mathbb P}\left(
\begin{tikzpicture}[baseline=8pt]
 \draw [fill] (0,0) circle(0.2ex);
 \draw [fill] (0.2,0) circle(0.2ex);
 \draw [fill] (0,0.85) circle(0.2ex);
 \draw [fill] (0.2,0.85) circle(0.2ex);
 \draw [thick] (0.2,0)--(0.2,0.85);
 \end{tikzpicture}
\right)}

\newcommand{\Pannb}{{\mathbb P}\left(
\begin{tikzpicture}[baseline=8pt]
 \draw [fill] (0,0) circle(0.2ex);
 \draw [fill] (0.2,0) circle(0.2ex);
 \draw [fill] (0,0.85) circle(0.2ex);
 \draw [fill] (0.2,0.85) circle(0.2ex);
 \draw [thick] (0,0)--(0.2,0.85);
 \end{tikzpicture}
\right)}

\newcommand{\Pnban}{{\mathbb P}\left(
\begin{tikzpicture}[baseline=8pt]
 \draw [fill] (0,0) circle(0.2ex);
 \draw [fill] (0.2,0) circle(0.2ex);
 \draw [fill] (0,0.85) circle(0.2ex);
 \draw [fill] (0.2,0.85) circle(0.2ex);
 \draw [thick] (0.2,0)--(0.,0.85);
 \end{tikzpicture}
\right)}

\newcommand{\Pabc}{{\mathbb P}\left(
\begin{tikzpicture}[baseline=8pt]
 \draw [fill] (0,0) circle(0.2ex);
 \draw [fill] (0.2,0) circle(0.2ex);
 \draw [fill] (0,0.85) circle(0.2ex);
 \draw [fill] (0.2,0.85) circle(0.2ex);
  \draw [fill] (0.4,0.) circle(0.2ex);
 \draw [fill] (0.4,0.85) circle(0.2ex);
 \draw [thick] (0,0)--(0,0.85);
 \draw [thick] (0.2,0)--(0.2,0.85);
  \draw [thick] (0.4,0)--(0.4,0.85);
 \end{tikzpicture}
\right)}

\newcommand{\Pbac}{{\mathbb P}\left(
\begin{tikzpicture}[baseline=8pt]
 \draw [fill] (0,0) circle(0.2ex);
 \draw [fill] (0.2,0) circle(0.2ex);
 \draw [fill] (0,0.85) circle(0.2ex);
 \draw [fill] (0.2,0.85) circle(0.2ex);
  \draw [fill] (0.4,0.85) circle(0.2ex);
   \draw [fill] (0.4,0) circle(0.2ex);
 \draw [thick] (0,0)--(0.2,0.85);
 \draw [thick] (0.2,0)--(0.,0.85);
  \draw [thick] (0.4,0)--(0.4,0.85);
 \end{tikzpicture}
\right)}

\newcommand{\Pacb}{{\mathbb P}\left(
\begin{tikzpicture}[baseline=8pt]
 \draw [fill] (0,0) circle(0.2ex);
 \draw [fill] (0.2,0) circle(0.2ex);
 \draw [fill] (0,0.85) circle(0.2ex);
 \draw [fill] (0.2,0.85) circle(0.2ex);
  \draw [fill] (0.4,0.85) circle(0.2ex);
   \draw [fill] (0.4,0) circle(0.2ex);
 \draw [thick] (0.2,0)--(0.4,0.85);
 \draw [thick] (0.4,0)--(0.2,0.85);
  \draw [thick] (0.,0)--(0.,0.85);
 \end{tikzpicture}
\right)}

\newcommand{\Pcba}{{\mathbb P}\left(
\begin{tikzpicture}[baseline=8pt]
 \draw [fill] (0,0) circle(0.2ex);
 \draw [fill] (0.2,0) circle(0.2ex);
 \draw [fill] (0,0.85) circle(0.2ex);
 \draw [fill] (0.2,0.85) circle(0.2ex);
  \draw [fill] (0.4,0.85) circle(0.2ex);
   \draw [fill] (0.4,0) circle(0.2ex);
 \draw [thick] (0.,0)--(0.4,0.85);
 \draw [thick] (0.4,0)--(0.,0.85);
  \draw [thick] (0.2,0)--(0.2,0.85);
 \end{tikzpicture}
\right)}

\newcommand{\Pbca}{{\mathbb P}\left(
\begin{tikzpicture}[baseline=8pt]
 \draw [fill] (0,0) circle(0.2ex);
 \draw [fill] (0.2,0) circle(0.2ex);
 \draw [fill] (0,0.85) circle(0.2ex);
 \draw [fill] (0.2,0.85) circle(0.2ex);
  \draw [fill] (0.4,0.85) circle(0.2ex);
   \draw [fill] (0.4,0) circle(0.2ex);
 \draw [thick] (0.,0)--(0.2,0.85);
  \draw [thick] (0.2,0.)--(0.4,0.85);
 \draw [thick] (0.4,0)--(0.,0.85);
 \end{tikzpicture}
\right)}

\newcommand{\Pcab}{{\mathbb P}\left(
\begin{tikzpicture}[baseline=8pt]
 \draw [fill] (0,0) circle(0.2ex);
 \draw [fill] (0.2,0) circle(0.2ex);
 \draw [fill] (0,0.85) circle(0.2ex);
 \draw [fill] (0.2,0.85) circle(0.2ex);
  \draw [fill] (0.4,0.85) circle(0.2ex);
   \draw [fill] (0.4,0) circle(0.2ex);
 \draw [thick] (0.,0)--(0.4,0.85);
  \draw [thick] (0.2,0.)--(0.,0.85);
 \draw [thick] (0.4,0)--(0.2,0.85);
 \end{tikzpicture}
\right)}

\newcommand{\Pex}{{\mathbb P}\left(
\begin{tikzpicture}[baseline=8pt]
 \draw [fill] (0,0) circle(0.2ex);
 \draw [fill] (0.2,0) circle(0.2ex);
 \draw [fill] (0,0.85) circle(0.2ex);
 \draw [fill] (0.2,0.85) circle(0.2ex);
  \draw [fill] (0.4,0.85) circle(0.2ex);
   \draw [fill] (0.4,0) circle(0.2ex);
  \draw [thick] (0.2,0.)--(0.,0.85);
 \draw [thick] (0.4,0)--(0.4,0.85);
 \end{tikzpicture}
\right)}

\newcommand{\Pabn}{{\mathbb P}\left(
\begin{tikzpicture}[baseline=8pt]
 \draw [fill] (0,0) circle(0.2ex);
 \draw [fill] (0.2,0) circle(0.2ex);
 \draw [fill] (0,0.85) circle(0.2ex);
 \draw [fill] (0.2,0.85) circle(0.2ex);
  \draw [fill] (0.4,0.85) circle(0.2ex);
   \draw [fill] (0.4,0) circle(0.2ex);
 \draw [thick] (0.,0)--(0.,0.85);
  \draw [thick] (0.2,0.)--(0.2,0.85);
 \end{tikzpicture}
\right)}

\newcommand{\Pabnabn}{{\mathbb P}\left(
\begin{tikzpicture}[baseline=8pt]
 \draw [fill] (0,0) circle(0.2ex);
 \draw [fill] (0.2,0) circle(0.2ex);
 \draw [fill] (0,0.85) circle(0.2ex);
 \draw [fill] (0.2,0.85) circle(0.2ex);
  \draw [fill] (0.4,0.85) circle(0.2ex);
   \draw [fill] (0.4,0) circle(0.2ex);
 \draw [thick] (0.,0)--(0.,0.85);
  \draw [thick] (0.2,0.)--(0.2,0.85);
 \end{tikzpicture}
\right)}

\newcommand{\Pabnanb}{{\mathbb P}\left(
\begin{tikzpicture}[baseline=8pt]
 \draw [fill] (0,0) circle(0.2ex);
 \draw [fill] (0.2,0) circle(0.2ex);
 \draw [fill] (0,0.85) circle(0.2ex);
 \draw [fill] (0.2,0.85) circle(0.2ex);
  \draw [fill] (0.4,0.85) circle(0.2ex);
   \draw [fill] (0.4,0) circle(0.2ex);
 \draw [thick] (0.,0)--(0.,0.85);
  \draw [thick] (0.2,0.)--(0.4,0.85);
 \end{tikzpicture}
\right)}

\newcommand{\Pabnnab}{{\mathbb P}\left(
\begin{tikzpicture}[baseline=8pt]
 \draw [fill] (0,0) circle(0.2ex);
 \draw [fill] (0.2,0) circle(0.2ex);
 \draw [fill] (0,0.85) circle(0.2ex);
 \draw [fill] (0.2,0.85) circle(0.2ex);
  \draw [fill] (0.4,0.85) circle(0.2ex);
   \draw [fill] (0.4,0) circle(0.2ex);
 \draw [thick] (0.,0)--(0.2,0.85);
  \draw [thick] (0.2,0.)--(0.4,0.85);
 \end{tikzpicture}
\right)}

\newcommand{\Pabnban}{{\mathbb P}\left(
\begin{tikzpicture}[baseline=8pt]
 \draw [fill] (0,0) circle(0.2ex);
 \draw [fill] (0.2,0) circle(0.2ex);
 \draw [fill] (0,0.85) circle(0.2ex);
 \draw [fill] (0.2,0.85) circle(0.2ex);
  \draw [fill] (0.4,0.85) circle(0.2ex);
   \draw [fill] (0.4,0) circle(0.2ex);
 \draw [thick] (0.,0)--(0.2,0.85);
  \draw [thick] (0.2,0.)--(0.,0.85);
 \end{tikzpicture}
\right)}

\newcommand{\Pabnbna}{{\mathbb P}\left(
\begin{tikzpicture}[baseline=8pt]
 \draw [fill] (0,0) circle(0.2ex);
 \draw [fill] (0.2,0) circle(0.2ex);
 \draw [fill] (0,0.85) circle(0.2ex);
 \draw [fill] (0.2,0.85) circle(0.2ex);
  \draw [fill] (0.4,0.85) circle(0.2ex);
   \draw [fill] (0.4,0) circle(0.2ex);
 \draw [thick] (0.,0)--(0.4,0.85);
  \draw [thick] (0.2,0.)--(0.,0.85);
 \end{tikzpicture}
\right)}

\newcommand{\Pabnnba}{{\mathbb P}\left(
\begin{tikzpicture}[baseline=8pt]
 \draw [fill] (0,0) circle(0.2ex);
 \draw [fill] (0.2,0) circle(0.2ex);
 \draw [fill] (0,0.85) circle(0.2ex);
 \draw [fill] (0.2,0.85) circle(0.2ex);
  \draw [fill] (0.4,0.85) circle(0.2ex);
   \draw [fill] (0.4,0) circle(0.2ex);
 \draw [thick] (0.,0)--(0.4,0.85);
  \draw [thick] (0.2,0.)--(0.2,0.85);
 \end{tikzpicture}
\right)}

\newcommand{\Panbabn}{{\mathbb P}\left(
\begin{tikzpicture}[baseline=8pt]
 \draw [fill] (0,0) circle(0.2ex);
 \draw [fill] (0.2,0) circle(0.2ex);
 \draw [fill] (0,0.85) circle(0.2ex);
 \draw [fill] (0.2,0.85) circle(0.2ex);
  \draw [fill] (0.4,0.85) circle(0.2ex);
   \draw [fill] (0.4,0) circle(0.2ex);
 \draw [thick] (0.,0)--(0.,0.85);
  \draw [thick] (0.4,0.)--(0.2,0.85);
 \end{tikzpicture}
\right)}

\newcommand{\Panbanb}{{\mathbb P}\left(
\begin{tikzpicture}[baseline=8pt]
 \draw [fill] (0,0) circle(0.2ex);
 \draw [fill] (0.2,0) circle(0.2ex);
 \draw [fill] (0,0.85) circle(0.2ex);
 \draw [fill] (0.2,0.85) circle(0.2ex);
  \draw [fill] (0.4,0.85) circle(0.2ex);
   \draw [fill] (0.4,0) circle(0.2ex);
 \draw [thick] (0.,0)--(0.,0.85);
\draw [thick] (0.4,0.)--(0.4,0.85);
 \end{tikzpicture}
\right)}

\newcommand{\Panbnab}{{\mathbb P}\left(
\begin{tikzpicture}[baseline=8pt]
 \draw [fill] (0,0) circle(0.2ex);
 \draw [fill] (0.2,0) circle(0.2ex);
 \draw [fill] (0,0.85) circle(0.2ex);
 \draw [fill] (0.2,0.85) circle(0.2ex);
  \draw [fill] (0.4,0.85) circle(0.2ex);
   \draw [fill] (0.4,0) circle(0.2ex);
 \draw [thick] (0.,0)--(0.2,0.85);
\draw [thick] (0.4,0.)--(0.4,0.85);
 \end{tikzpicture}
\right)}

\newcommand{\Panbban}{{\mathbb P}\left(
\begin{tikzpicture}[baseline=8pt]
 \draw [fill] (0,0) circle(0.2ex);
 \draw [fill] (0.2,0) circle(0.2ex);
 \draw [fill] (0,0.85) circle(0.2ex);
 \draw [fill] (0.2,0.85) circle(0.2ex);
  \draw [fill] (0.4,0.85) circle(0.2ex);
   \draw [fill] (0.4,0) circle(0.2ex);
 \draw [thick] (0.,0)--(0.2,0.85);
\draw [thick] (0.4,0.)--(0.,0.85);
 \end{tikzpicture}
\right)}

\newcommand{\Panbbna}{{\mathbb P}\left(
\begin{tikzpicture}[baseline=8pt]
 \draw [fill] (0,0) circle(0.2ex);
 \draw [fill] (0.2,0) circle(0.2ex);
 \draw [fill] (0,0.85) circle(0.2ex);
 \draw [fill] (0.2,0.85) circle(0.2ex);
  \draw [fill] (0.4,0.85) circle(0.2ex);
   \draw [fill] (0.4,0) circle(0.2ex);
 \draw [thick] (0.,0)--(0.4,0.85);
\draw [thick] (0.4,0.)--(0.,0.85);
 \end{tikzpicture}
\right)}

\newcommand{\Panbnba}{{\mathbb P}\left(
\begin{tikzpicture}[baseline=8pt]
 \draw [fill] (0,0) circle(0.2ex);
 \draw [fill] (0.2,0) circle(0.2ex);
 \draw [fill] (0,0.85) circle(0.2ex);
 \draw [fill] (0.2,0.85) circle(0.2ex);
  \draw [fill] (0.4,0.85) circle(0.2ex);
   \draw [fill] (0.4,0) circle(0.2ex);
 \draw [thick] (0.,0)--(0.4,0.85);
\draw [thick] (0.4,0.)--(0.2,0.85);
 \end{tikzpicture}
\right)}

\newcommand{\Pnababn}{{\mathbb P}\left(
\begin{tikzpicture}[baseline=8pt]
 \draw [fill] (0,0) circle(0.2ex);
 \draw [fill] (0.2,0) circle(0.2ex);
 \draw [fill] (0,0.85) circle(0.2ex);
 \draw [fill] (0.2,0.85) circle(0.2ex);
  \draw [fill] (0.4,0.85) circle(0.2ex);
   \draw [fill] (0.4,0) circle(0.2ex);
 \draw [thick] (0.2,0)--(0.,0.85);
  \draw [thick] (0.4,0.)--(0.2,0.85);
 \end{tikzpicture}
\right)}

\newcommand{\Pnabanb}{{\mathbb P}\left(
\begin{tikzpicture}[baseline=8pt]
 \draw [fill] (0,0) circle(0.2ex);
 \draw [fill] (0.2,0) circle(0.2ex);
 \draw [fill] (0,0.85) circle(0.2ex);
 \draw [fill] (0.2,0.85) circle(0.2ex);
  \draw [fill] (0.4,0.85) circle(0.2ex);
   \draw [fill] (0.4,0) circle(0.2ex);
 \draw [thick] (0.2,0)--(0.,0.85);
\draw [thick] (0.4,0.)--(0.4,0.85);
 \end{tikzpicture}
\right)}

\newcommand{\Pnabnab}{{\mathbb P}\left(
\begin{tikzpicture}[baseline=8pt]
 \draw [fill] (0,0) circle(0.2ex);
 \draw [fill] (0.2,0) circle(0.2ex);
 \draw [fill] (0,0.85) circle(0.2ex);
 \draw [fill] (0.2,0.85) circle(0.2ex);
  \draw [fill] (0.4,0.85) circle(0.2ex);
   \draw [fill] (0.4,0) circle(0.2ex);
 \draw [thick] (0.2,0)--(0.2,0.85);
\draw [thick] (0.4,0.)--(0.4,0.85);
 \end{tikzpicture}
\right)}

\newcommand{\Pnabban}{{\mathbb P}\left(
\begin{tikzpicture}[baseline=8pt]
 \draw [fill] (0,0) circle(0.2ex);
 \draw [fill] (0.2,0) circle(0.2ex);
 \draw [fill] (0,0.85) circle(0.2ex);
 \draw [fill] (0.2,0.85) circle(0.2ex);
  \draw [fill] (0.4,0.85) circle(0.2ex);
   \draw [fill] (0.4,0) circle(0.2ex);
 \draw [thick] (0.2,0)--(0.2,0.85);
\draw [thick] (0.4,0.)--(0.,0.85);
 \end{tikzpicture}
\right)}

\newcommand{\Pnabbna}{{\mathbb P}\left(
\begin{tikzpicture}[baseline=8pt]
 \draw [fill] (0,0) circle(0.2ex);
 \draw [fill] (0.2,0) circle(0.2ex);
 \draw [fill] (0,0.85) circle(0.2ex);
 \draw [fill] (0.2,0.85) circle(0.2ex);
  \draw [fill] (0.4,0.85) circle(0.2ex);
   \draw [fill] (0.4,0) circle(0.2ex);
 \draw [thick] (0.2,0)--(0.4,0.85);
\draw [thick] (0.4,0.)--(0.,0.85);
 \end{tikzpicture}
\right)}

\newcommand{\Pnabnba}{{\mathbb P}\left(
\begin{tikzpicture}[baseline=8pt]
 \draw [fill] (0,0) circle(0.2ex);
 \draw [fill] (0.2,0) circle(0.2ex);
 \draw [fill] (0,0.85) circle(0.2ex);
 \draw [fill] (0.2,0.85) circle(0.2ex);
  \draw [fill] (0.4,0.85) circle(0.2ex);
   \draw [fill] (0.4,0) circle(0.2ex);
 \draw [thick] (0.2,0)--(0.4,0.85);
\draw [thick] (0.4,0.)--(0.2,0.85);
 \end{tikzpicture}
\right)}

\newcommand{\Pannann}{{\mathbb P}\left(
\begin{tikzpicture}[baseline=8pt]
 \draw [fill] (0,0) circle(0.2ex);
 \draw [fill] (0.2,0) circle(0.2ex);
 \draw [fill] (0,0.85) circle(0.2ex);
 \draw [fill] (0.2,0.85) circle(0.2ex);
  \draw [fill] (0.4,0.85) circle(0.2ex);
   \draw [fill] (0.4,0) circle(0.2ex);
 \draw [thick] (0.,0)--(0.,0.85);
 \end{tikzpicture}
\right)}

\newcommand{\Pannnna}{{\mathbb P}\left(
\begin{tikzpicture}[baseline=8pt]
 \draw [fill] (0,0) circle(0.2ex);
 \draw [fill] (0.2,0) circle(0.2ex);
 \draw [fill] (0,0.85) circle(0.2ex);
 \draw [fill] (0.2,0.85) circle(0.2ex);
  \draw [fill] (0.4,0.85) circle(0.2ex);
   \draw [fill] (0.4,0) circle(0.2ex);
 \draw [thick] (0.,0)--(0.4,0.85);
 \end{tikzpicture}
\right)}

\newcommand{\Pnnaann}{{\mathbb P}\left(
\begin{tikzpicture}[baseline=8pt]
 \draw [fill] (0,0) circle(0.2ex);
 \draw [fill] (0.2,0) circle(0.2ex);
 \draw [fill] (0,0.85) circle(0.2ex);
 \draw [fill] (0.2,0.85) circle(0.2ex);
  \draw [fill] (0.4,0.85) circle(0.2ex);
   \draw [fill] (0.4,0) circle(0.2ex);
 \draw [thick] (0.4,0)--(0.,0.85);
 \end{tikzpicture}
\right)}

\newcommand{\Pnnanna}{{\mathbb P}\left(
\begin{tikzpicture}[baseline=8pt]
 \draw [fill] (0,0) circle(0.2ex);
 \draw [fill] (0.2,0) circle(0.2ex);
 \draw [fill] (0,0.85) circle(0.2ex);
 \draw [fill] (0.2,0.85) circle(0.2ex);
  \draw [fill] (0.4,0.85) circle(0.2ex);
   \draw [fill] (0.4,0) circle(0.2ex);
 \draw [thick] (0.4,0)--(0.4,0.85);
 \end{tikzpicture}
\right)}

\newcommand{\Pcbn}{{\mathbb P}\left(
\begin{tikzpicture}[baseline=8pt]
 \draw [fill] (0,0) circle(0.2ex);
 \draw [fill] (0.2,0) circle(0.2ex);
 \draw [fill] (0,0.85) circle(0.2ex);
 \draw [fill] (0.2,0.85) circle(0.2ex);
  \draw [fill] (0.4,0.85) circle(0.2ex);
   \draw [fill] (0.4,0) circle(0.2ex);
 \draw [thick] (0.0,0)--(0.4,0.85);
  \draw [thick] (0.2,0.)--(0.2,0.85);
 \end{tikzpicture}
\right)}

\usepackage{pstricks}

\usepackage{tikz}
\usetikzlibrary{calc}
\usetikzlibrary{chains}
\usetikzlibrary{scopes}
\usetikzlibrary{shapes.geometric}
\usetikzlibrary{trees}
\usetikzlibrary{topaths}
\usetikzlibrary{positioning}
\usetikzlibrary{patterns,decorations.pathreplacing}

\begin{document}

\title[Non-scalar operators for the Potts model in arbitrary dimension]{Non-scalar operators for the Potts model in arbitrary dimension}
\author{Romain Couvreur$^{1,2,3}$, Jesper Lykke Jacobsen$^{1,2,3}$ and Romain Vasseur$^{4,5}$}

\address{${}^1$Institut de Physique Th\'eorique, CEA Saclay, 91191 
Gif Sur Yvette, France}
\address{${}^2$Laboratoire de physique th\'eorique, D\'epartement de physique de l'ENS, \'Ecole Normale Sup\'erieure,
UPMC Univ. Paris 06, CNRS, PSL Research University, 75005 Paris, France}
\address{${}^3$Sorbonne Universit\'es, UPMC Univ. Paris 06, \'Ecole Normale Sup\'erieure, CNRS, Laboratoire de Physique Th\'eorique (LPT ENS), 75005 Paris, France} 
\address{${}^4$Department of Physics, University of California, Berkeley, Berkeley CA 94720, USA}
\address{${}^5$Materials Science Division, Lawrence Berkeley National Laboratory, Berkeley CA 94720, USA}

\eads{\mailto{romain.couvreur@ens.fr}, 
      \mailto{jesper.jacobsen@ens.fr}, 
      \mailto{rvasseur@berkeley.edu}}

\begin{abstract}
We investigate the operator content of the $Q$-state Potts model in arbitrary dimension, using the representation theory of the symmetric group.
In particular we construct all possible tensors acting on $N$ spins, corresponding to given symmetries under $\Sc_Q$ and $\Sc_N$, in terms of representations
involving any Young diagram. These operators transform non-trivially under the group of spatial rotations, with a definite conformal spin.
The two-point correlation functions are then computed, and their physical interpretation is given in terms of Fortuin-Kasteleyn clusters propagating between
two neighbourhoods of each $N$ spins. In two dimensions, we obtain analytically the critical exponent corresponding to each operator. In the simplest and
physically most relevant cases, we confirm the values of the critical exponent and the conformal spin by numerical measurements, using both
Monte Carlo simulations and transfer matrix diagonalisations. 
Our classification partially provides the structure of Jordan cells of the dilatation operator in arbitrary dimensions, which in turn gives rise to logarithmic correlation functions. 
\end{abstract}




\section{Introduction}
Critical systems are well known for having correlation functions with power-law behavior. Conformal field theory (CFT) has been very successful in two dimensions, where a large class of scale invariant theories---the minimal models---have been classified \cite{BPZ,FQS,CardyMin}. The most widely used minimal models share a very important constraint: unitarity. In many interesting systems that come from statistical physics, unitarity is not verified. Geometrical models, such as percolation and the self-avoiding walk \cite{Saleur87}, or disordered systems, such as the plateau transition in the quantum Hall effect(s), are non unitary. These non-unitary conformal field theories contain another class of theories corresponding to the Logarithmic Conformal Field Theories (LCFT). They are special in the sense that the scale invariant correlation functions may contain logarithmic factors \cite{RS92,Gurarie93} as well as power laws. LCFTs appear in an extended version of minimal models where some operators describe non-local geometrical correlations \cite{PRZ06,RS07}.

LCFTs are characterised by having a non-diagonalisable dilatation operator $L_0$. This structure leads to logarithms in the correlations of the fields that transform inside a Jordan cell of $L_0$. Since the above pioneering works it has become a very active subject, where much progress was made thanks to the study of indecomposable algebras. However it was realised only later that some features of the LCFTs can be understood by taking a limiting procedure \cite{C99,LCFT2,VJS2012} through a continuous family of ordinary (i.e., not indecomposable, but still non-unitary) CFT. If the model is subject to additional discrete symmetries, insight on these logarithmic correlation functions is obtained by studying operators that transform irreducibly under this symmetry. In contradistinction to most of usual techniques in CFT, this approach is {\em not} limited to two dimensions.

The $Q$-state Potts model is an interesting statistical physics system, since it contains several important special cases, such as the Ising model ($Q=2$), percolation ($Q \to 1$), or spanning trees and forests ($Q \to 0$ limits) \cite{JSS05,CJSSS04,JS05,DGS07,CSS17}. This model is invariant under the action of the discrete symmetric group $\Sc_Q$. Even though the original definition supposes $Q$ to be a positive integer, the partition function can be reformulated in terms of non-local geometrical objects and analytically continued to $Q$ real \cite{FK}. More details can be found in section~\ref{Potts} below. In two dimensions, the Potts model can be solved exactly thanks to a mapping to a 6-vertex model \cite{BKW76} or an associated loop model \cite{TL71}. Very little is known about the field theory in higher dimensions. A first approach to the logarithmic scale invariant theory for the Potts model has been proposed in \cite{VJS2012}, and subsequently investigated more systematically by two of the present authors \cite{VJ2014}. It successfully predicts Jordan cells in the dilatation operator, independently of the dimension. Recently, progress has been made to determine the general form of correlation functions and conformal blocks in any dimension \cite{HPV16}. 

In this paper we generalise the construction of \cite{VJ2014} to classify local operators acting on $N$ spins that transform irreducibly under the action of the symmetric group, taking into account the $\Sc_Q$ symmetry and also the $\Sc_N$ symmetry that describes the behaviour of the operators under an exchange of the spins. Only operators symmetric under such exchange were previously considered in \cite{VJ2014}, corresponding to a trivial action by $\Sc_N$ and a zero conformal spin. Here we consider arbitrary irreducible operators, including those with non-zero conformal spin, in a more transparent group theoretical setup. The structure and the correlation functions of this new class of observables is discussed. In particular, we underline the consequences for bulk LCFT and identify the critical exponents in two dimensions.

\section{Operators of the $Q$-state Potts model}
\label{Potts}
The $Q$-state Potts model is a quite simple model of statistical physics with $\Sc_Q$ symmetry, which nevertheless conceals deep algebraic structures and a rich variety of critical behaviour. It can be written in different representations, some of which are specific to two dimensions, and others which hold in arbitrary dimensions. Among the latter, a very appealing choice amounts to rewriting the partition function and correlation functions in terms of non-local objects known as Fortuin-Kasteleyn clusters \cite{FK}. Each such cluster carries a Boltzmann weight of $Q$, which immediately allows us to extend the definition of the model to non-integer $Q$. This is particularly useful in the LCFT context, because we can then access particular indecomposable values of $Q$ through a limiting procedure.

In the following, we first provide a brief review of the definition of the Potts model. We then explain the construction of local operators within a particular representation of $\Sc_Q$ and discuss the structure relevant to LCFT. The previous results for scalar (spinless) operators \cite{VJ2014} are recovered within our new framework.

 \subsection{Potts model and Fortuin-Kasteleyn clusters}
\label{sec:Potts-FK}

The $Q$-state Potts model is a lattice model of interacting spins. Considering a graph $G = (V,E)$ with vertex set $V$ and edge set $E$. To each vertex $i\in V$ we associate a spin $\sg_i$ that can take $Q$ possible states, $\sg_i = 1,2,\ldots,Q$. The interaction between two spins, $\sg_i$ and $\sg_j$, linked by an edge $(ij) \in E$, adds a contribution $-K\dt_{\sg_i,\sg_j}$ to the total energy of the system, where $\dt$ is the Kronecker symbol. Hence the partition function is
\be Z=\sum_{\{\sg\}}\prod_{(i,j)\in E}\ex^{K\dt_{\sg_i,\sg_j}} \,, \ee
where $\{\sg\} \equiv \{\sg_i\}_{i \in V}$ denotes all the possible spin configurations of the system. Let us emphasise the fact that for now we do not make any particular assumptions about the graph $G$, neither about the regularity of the lattice, nor about the number of dimensions $d$ into which it can be embedded.

The above partition function can be expanded in terms of the Fortuin-Kasteleyn (FK)~clusters. Since the interaction energy between two spins has only two possible values, we use the identity $\ex^{K\dt_{\sg_i,\sg_j}}=1+v\dt_{\sg_i,\sg_j}$, with $v=\ex^K-1$, to transform the product over the edges into a sum over subsets $A\subseteq E$ of edges. A connected component of the subgraph $(V,A)$ is called an FK cluster. Then, by performing the sum over $\{\sg\}$, we arrive at
\be Z=\sum_{A\subseteq E}Q^{k(A)}v^{|A|} \,, \label{Z-FK} \ee
where $k(A)$ denotes the number of connected components in the subgraph $(V,A)$ with $|A|$ edges. 

We notice that all the spins within the same FK cluster take the same spin state. However, the spins of two different clusters are independent; in particular they may take the same value even if the two clusters are adjacent.

Thanks to the formulation (\ref{Z-FK}) of the $Q$-state Potts model, the definition of the partition function can now be extended to real values of $Q$. This makes it possible to approach physical situations (for which $Q$ is a non-negative integer) via a limiting procedure. As mentioned in the introduction, the models of main interest stand at $Q=0$ (spanning trees and forests), $Q=1$ (percolation) and $Q=2$ (Ising model). For $Q<2$ they have non-trivial critical points for $2\leq d\leq 6$ \cite{DGS07,A76}. In $d=2$, the model on a regular square lattice can be mapped to a 6-vertex model \cite{BKW76} and is critical for $0\leq Q\leq4$ \cite{Baxter73}.

We are going to study the so-called $N$-cluster operators.%
\footnote{In $d=2$, the $N$-cluster operators coincide with the $2N$-leg watermelon (or fuseau) operators in the loop model representation. However, this latter representation makes crucial use of planarity and duality, and hence is not available in higher dimensions.}
Consider a small neighbourhood $\mathcal{D}_i$ containing $N$ spins that belong to $N$ distinct FK~clusters. We are interested in the probability that these $N$ clusters extend to another similar neighbourhood $\mathcal{D}_j$. A {\em cluster-operator} ${\sf O}_i$---which will be defined precisely below in a more appropriate notation---acts on the $N$ spins in $\mathcal{D}_i$ so as to impose that they belong to {\em distinct} FK~clusters. The clusters thus inserted at $\mathcal{D}_i$ are therefore obliged to propagate until they terminate at another cluster operator.%
\footnote{The use of words such as ``insert'', ``propagate'' and ``terminate'' does not imply that these clusters have been equipped with a direction of any kind.}
The probability of having $N$ distinct clusters extending from $\mathcal{D}_i$ to $\mathcal{D}_j$ can thus be related to the two-point function $\langle {\sf O}_i {\sf O}_j \rangle$.

We shall see below that the $N$-cluster operator ${\sf O}_i$ acts as a linear operator on the $N$ Potts spins in $\mathcal{D}_i$. It can therefore be represented explicitly as an $N$-component tensor (e.g., a vector for $N=1$ and a matrix for $N=2$) of coefficients that are algebraic functions of $Q$.
For that reason we shall often use the words operator and tensor interchangingly in the sequel.

The correlation functions in the Potts model at criticality involve power laws of the distance between the small neighbourhoods considered. If $Q$ takes the particular values $Q=4\cos^2(\pi/p)$ with $p=2,3,\ldots$ integer (these values are known as Beraha numbers) in $d=2$, or $Q\in\mathbb{N}$ in higher dimensions, we will find that these power laws are multiplied by logarithms. In two dimensions, these logarithmic contributions have been fixed \cite{Gurarie93}.
In \cite{VJ2014}, two of the present authors have constructed all the tensors acting on an arbitrary number of spins and transforming according to a class of irreducible representations of the symmetric group. More precisely, they have computed the form of all possible tensors $t(\sg_1,\dots,\sg_N)$, acting on $N$ spins, which are irreducible under $\Sc_Q$ and invariant under arbitrary permutations of the $N$ spins $\sg_i$.

\subsection{Definitions and representation theory of $\Sc_Q$}
The symmetric group $\Sc_n$ is the group of permutations of $n$ integers. We would like to classify the operator content of the Potts model in terms of representation of $\Sc_Q$. We need to construct an operator acting on $N$ spins, $t(\sg_1,\ldots,\sg_N)$. Let us denote by $L_Q^{(N)}$ the space of $\underbrace{Q\times...\times Q}_{N}$ tensors. We decompose it in terms of the basis elements $\Op_{a_1,a_2,\ldots,a_N}$ defined by
\be \Op_{a_1,a_2,\ldots,a_N}(\sg_1,\sg_2,\ldots,\sg_N)=\delta_{a_1,\sg_1}\delta_{a_2,\sg_2}\ldots\delta_{a_N,\sg_N} \,, \label{basis-elements} \ee
where $\{a_i\}$ and $\{\sg_i\}$ are sets of integers between $1$ and $Q$.
A permutation $p \in \Sc_Q$ has a natural action on this basis:
\be p\ \Op_{a_1,a_2,\ldots,a_N}=\Op_{p(a_1),p(a_2),\ldots,p(a_N)} \,. \ee
We can also define the action of $\sn{p}$, a permutation in $\Sc_N$, by
\bea\label{snaction}\sn{p}\ \Op_{a_1,a_2,\ldots,a_N}&=&\Op_{a_{\sn{p}^{-1}(1)},a_{\sn{p}^{-1}(2)},\ldots,a_{\sn{p}^{-1}(N)}} \,,\eea
which implies:
\bea\sn{p}\ \Op_{a_1,a_2,\ldots,a_N}(\sg_1,\ldots,\sg_N)&=&\Op_{a_1,a_2,\ldots,a_N}(\sg_{\sn{p}(1)},\ldots,\sg_{\sn{p}(N)}) \,.\eea

It is straightforward to show that if we act with $p\in\Sc_Q$ and $\sn{p}\in\Sc_N$ on a tensor, the order does not matter because the transformations commute.

To make the presentation self-contained, we start by briefly recalling all the necessary definitions and properties. We refer to \cite{WKT} for more details. In the following, a permutation is always represented by its decomposition in terms of cycles. A very convenient way to study the representations of $\Sc_n$ is to use the bijection between irreducible representations and Young diagrams. Let us consider a partition of the integer $n$, with $k$ integers $\la_i$ ($\sum_{i=1}^k\la_i=n$) that we reorder such that $\la_1\geq\la_2\geq\cdots\geq\la_k$. We use the standard notation for Young diagrams, where $\left[\la_1,\la_2,\ldots\right]$ denotes the diagram with $\la_i$ boxes in the $i$'th row (counted from the top). The normal Young tableau corresponding to a Young diagram is obtained by filling the $n$ boxes with the integers $\{1,2,\ldots,n\}$, so that the order is increasing in each row (read from left to right) and in each column (read from top to bottom). We henceforth refer to rows also as horizontal lines, and to columns as vertical lines. An arbitrary Young tableau is obtained by acting with a permutation $q$ on the numbers in the boxes of the normal tableau. Therefore the quantities that refer to a particular Young tableau of shape $\lambda$ will be denoted with a superscript $(q)$.

We now define a linear operator which will project a tensor in a certain irreducible representation of the symmetric group. Given a Young tableau $\la$ of size $n$ boxes, let $h_{\la}^{(q)}$ (resp.\ $v_{\la}^{(q)}$) be the subset of permutations that leave invariant the set of numbers in each horizontal (resp.\ vertical) line of the Young tableaux of shape $\la$ and filled thanks to the permutation $q$. We then define the symmetriser $s_{\la}^{(q)}$, the anti-symmetriser $a_{\la}^{(q)}$ and the irreducible symmetriser $e_{\la}^{(q)}$ by
\be s_{\la}^{(q)}=\sum_{h\in h_{\la}^{(q)}}h \,, \qquad a_{\la}^{(q)}=\sum_{v\in v_{\la}^{(q)}}\epsilon(v)v \,, \qquad e_{\la}^{(q)}=s_\la^{(q)} a_{\la}^{(q)} \,, \label{irreduc-symm} \ee
where $\epsilon(v)$ is the signature of the permutation $v$.
The idea is the following: when we act with the operator $e_{\la}^{(q)}$ on a set of indices, first all the indices in the same column are anti-symmetrised, and then we symmetrise all the indices in the same row. Notice that the quantity that we obtain is a priori not anti-symmetric in the indices that belong to the same column, since $s_\la^{(q)}$ and $a_\la^{(q)}$ do not commute. Examples of irreducible symmetrisers will be given when we will construct the tensors. Finally, notice that
\be e_\la^{(q)}=qe_\la q^{-1}\label{changesym} \,, \ee
which is a relation that relates the irreducible symmetriser of any Young tableau of shape $\la$ to the irreducible symmetriser of the normal Young tableau of the same shape.
The operator $e_{\la}^{(q)}$ is a projector (if normalised correctly). In order to get an invariant subset of a vector space (which will be our operator space) that generates an irreducible representation, we act with $e_{\la}^{(q)}$ on an arbitrary vector. All the other generators are obtained by acting with permutations on this first generator. A very important fact is that if one changes the order of the numbers appearing in the boxes of the diagram $\la$, we end up with an equivalent representation. Later we will choose those numbers conveniently.

Let us finally recall the {\em hook formula} that gives the dimension $d_\la$ of an irreducible representation $\la$ of $\Sc_Q$:
\be
 d_\la = \frac{n!}{ \prod_{(i,j) \in \lambda} h_\la(i,j)} \,, \label{hook-formula}
\ee
where the hook length $h_\la(i,j)$ of the box labelled $(i,j)$ in the Young diagram $\lambda$ is the number of boxes to its right, plus the number of boxes below it, plus the box itself.

We are now ready to act with an irreducible symmetriser on this vector space to obtain a family of tensors in an arbitrary representation of $\Sc_Q$. Let us summarise the ``recipe''. First we choose a Young diagram $\la_Q$ of size $Q$. We consider a realisation of a Young tableau filled with the permutation $q$ and compute the irreducible symmetriser $e_{\la_Q}^{(q)}$. The action of this operator on a tensor $t$ gives us a generator $e_{\la_Q}^{(q)}t$ of an invariant subspace $V_{\la_Q}$. We get all the generators of $V_{\la_Q}$ that transform with the $\Sc_Q$ irreducible representation associated to $\la_Q$ by acting with some permutations:
\be V_{\la_Q}= \left \{ p\left(e_{\la_Q}^{(q)}t\right) \, : \, p\in\Sc_Q \right \} \,. \ee

In the following, we review the simplest examples of this procedure, recovering some of the results of \cite{VJ2014}, before generalising the method.

\subsection{Operators acting on one spin}
First we consider the vector space $L_Q^{(1)}=\left\{\Op_a \,: a=1,\ldots,Q\right\}$ of operators acting on one spin $\sigma$. In \cite{VJS2012,VJ2014}, the authors decomposed this space in two operators, namely the identity and the magnetisation. In their notation, these operators read
\bea t^{(0,1)}(\sg)&=&\sum_{a=1}^Q\dt_{a,\sg} =1 \,, \label{identity1}\\
t_a^{(1,1)}(\sg)&=&\dt_{a,\sg}-\tfrac{1}{Q}t^{(0,1)}=\dt_{a,\sg}-\tfrac{1}{Q} \,. \label{magnetization1}\eea
These expressions were found in \cite{VJS2012,VJ2014} by imposing the constraint
\be \sum_at_a^{(1,1)}=0 \,. \label{magnetization-constraint} \ee
Below we shall rederive them (in our own notation), using a procedure that allows us to consider more complicated cases later. To this end, we consider the irreducible symmetriser $e^{(q)}_\lambda$ for a given Young tableau $\lambda$. It is easy to see that only two Young diagrams give a non-zero action: $[Q]$ and $[Q-1,1]$. These diagrams are respectively associated to the trivial and standard representations. Their respective dimensions ($1$ and $Q-1$), as found from (\ref{hook-formula}), justifies the constraint (\ref{magnetization-constraint}).

\subsubsection{Invariant tensor.}
The irreducible symmetriser for $[Q]$ is
\be e_{[Q]} = \sum_{h\in\Sc_Q}h \,. \ee
Its action on any tensor $\Op_a$ gives the invariant tensor $t^{[Q]}$:
\be t^{[Q]}\equiv\tfrac{1}{(Q-1)!}e_{[Q]}\Op_a = \sum_{i=1}^Q\Op_i \,. \ee
The tensor acting on a spin $\sg$ is explicitly
\be t^{[Q]}(\sg)=1 \,, \ee
and so we have recovered \eqref{identity1}.

\subsubsection{Standard representation.}
Let us now move to a more interesting case with the Young diagram $[Q-1,1]$ and consider the following Young tableau
\be
\ytableausetup{boxsize=2em,centertableaux}
\begin{ytableau}
 i_1  & i_2  & \dots  & \, i_{Q-1} \\
 a
\end{ytableau} \label{Q-1-YT}
\ee
with $a$ a specific index (with $a=1,\ldots,Q$) and $i_1,\ldots,i_{Q-1}$ arbitrary indices in order to fill the Young tableau.
The particular choice $\{ i_k\}=\{1,2,\ldots,a-1,a+1,\ldots,Q\}$ gives the normal Young tableau for the given value of $a$.
We will also denote by $a$ the permutation that produces the particular tableau (\ref{Q-1-YT}) from the corresponding normal Young tableau.
The corresponding irreducible symmetriser $e_{\la_Q}^{(a)}$ is
\be e_{\la_Q}^{(a)}=\left(\sum_{h\in \tilde{h}_\la^{(a)}}h\right)\left(1-(i_1,a)\right) \,, \ee
where $\tilde{h}_\la^{(a)}$ is the subset of permutations within $\Sc_Q$ that leave invariant the first line of the Young tableau or, in other words, the permutations leaving $a$ invariant. Also, $(i_1,a)$ denotes the transposition of the indices $i_1$ and $a$. Acting on the basis element $\Op_a$ we get
\bea e_{\la_Q}^{(a)} \Op_a &=& \left(\sum_{h\in \tilde{h}_\la^{(a)}}h\right)\left(\Op_a-\Op_{i_1}\right) \nonumber \\
&=&(Q-1)!\Op_a-(Q-2)!\sum_{\substack{i=1\\ i\neq a}}^{Q}\Op_i \nonumber \\
&=&Q(Q-2)!\Op_a-(Q-2)!\sum_{i=1}^{Q}\Op_i \,. \eea
We define the normalised tensor $t^{[Q-1,1]}_a$ by
\be \label{tensor1} t^{[Q-1,1]}_a=\tfrac{1}{Q(Q-2)!}e_{\la_Q}^{(a)} \Op_a=\Op_a-\tfrac{1}{Q}\sum_{i=1}^{Q}\Op_i=\Op_a-\tfrac{1}{Q}t^{[Q]} \,. \ee
We can generate the whole family of tensors by acting with a permutation $p\in\Sc_Q$, but we find that this family is just given by the previous formula for any $a$, because of (\ref{changesym}). We can verify that it has dimension $Q-1$, because of the following identity
\be \sum_{a=1}^Q t^{[Q-1,1]}_a=0 \,,\ee
which of course is nothing but (\ref{magnetization-constraint}) in our notation.
The tensor acting on a spin $\sg$ is explicitly
\be t^{[Q-1,1]}_a(\sg)=\dt_{a,\sg}-\tfrac{1}{Q} \,, \ee
so we have recovered also the magnetisation operator \eqref{magnetization1}.

Summarising, the space $L_Q^{(1)}$ has be decomposed as a direct sum of two irreducible representations of $\Sc_Q$: 
\be L_Q^{(1)}=[Q]\oplus[Q-1,1] \,.\ee

\subsection{Operators acting on $2$ spins}
We next consider the space $L_Q^{(2)}$ of operators acting on $2$ spins, $\sigma_1$ and $\sigma_2$.
The subspace of $L_Q^{(2)}$ of operators for which $\sigma_1 = \sigma_2$ reads $\{\Op_{a,a} \,: a=1,\ldots,Q\}$ and is stable under the action of $\Sc_Q$. It is isomorphic to $L_Q^{(1)}$ and can therefore be decomposed as the direct sum $[Q]\oplus[Q-1,1]$, as we have just seen. We thus consider henceforth only operators for which $\sg_1\neq\sg_2$.

In \cite{VJ2014} the authors studied only the case of symmetric tensors, i.e., those satisfying $t(\sg_1,\sg_2)=t(\sg_2,\sg_1)$. They enforced constraints on the sum of coefficients to find three irreducible tensors, which we first discuss using the notation of \cite{VJ2014}. The first one, $t^{(0,2)}$, can be identified with the energy operator.
The second one, $t_a^{(1,2)}$, has the same $\Sc_Q$ symmetry as the magnetisation operator.
And the last tensor, $t_{a,b}^{(2,2)}$, is the two-cluster operator. The three tensors are found to have the following expressions \cite{VJS2012,VJ2014}:
\bea t^{(0,2)}(\sg_1,\sg_2)&=&1-\dt_{\sg_1,\sg_2} \,, \label{energy2}\\
t^{(1,2)}_a(\sg_1,\sg_2)&=&t_a^{(1,1)}(\sg_1)+t_a^{(1,1)}(\sg_2)=\dt_{a,\sg_1}+\dt_{a,\sg_2}-\tfrac{2}{Q} \,, \label{magnetization2}\\
t^{(2,2)}_{a,b}(\sg_1,\sg_2)&=&\dt_{a,\sg_1}\dt_{b,\sg_2}+\dt_{a,\sg_2}\dt_{b,\sg_1} \nonumber \\
& & -\tfrac{1}{Q-2}\left(t_a^{(1,2)}(\sg_1,\sg_2)+t_b^{(1,2)}(\sg_1,\sg_2)\right)-\tfrac{2}{Q(Q-1)}t^{(0,2)} \,, \label{2cluster2}\eea
where we have omitted writing an overall factor $(1-\delta_{\sg_1,\sg_2})$ multiplying the last two expressions.

As in the previous section we now show how to recover these expression using Young projectors. We also extend the set of operators by considering tensors which are antisymmetric under the permutation of the two spins, i.e., those satisfying $t(\sigma_1,\sigma_2) = -t(\sigma_2,\sigma_1)$. To this end, we first notice that any projector on a representation associated with a Young diagram with more than two boxes below the first line is the null operator. We thus have to consider the following diagrams: $[Q]$, $[Q-1,1]$, $[Q-2,2]$ and $[Q-2,1,1]$.

\subsubsection{Invariant tensor.}
\label{sec:energy-operator}
Let us again first consider the case of $[Q]$. The space $L_Q^{(2)}$ contains an additional invariant operator. When acting with $e_{[Q]}$ on the tensor $\Op_{a,b}$ with $a\neq b$ we find:
\be t^{[Q]} \equiv \tfrac{1}{(Q-2)!}e_{[Q]}\Op_{a,b} = \sum_{\substack{i,j=1\\i\neq j}}^Q\Op_{i,j} \,. \label{invariant2}\ee
In terms of the spins, $\sg_1$ and $\sg_2$, we have the expression
\be t^{[Q]}(\sg_1,\sg_2)= 1 - \delta_{\sg_1,\sg_2} \,, \ee
and so we have recovered \eqref{energy2} in our own notation.

\subsubsection{Representation $[Q-1,1]$.}
We now turn to the more interesting case of the Young diagram $[Q-1,1]$. We consider once again the following Young tableau:
\be
\ytableausetup{boxsize=2em,centertableaux}
\begin{ytableau}
 i_1 & i_2 & \ldots & \, i_{Q-1} \\
 a 
\end{ytableau}
\ee
One could act with the same $e_{\la_Q}^{(a)}$ as before on an element of the type $\Op_{a,b}$:
\bea
e_{\la_Q}^{(a)}\Op_{a,i_{Q-1}}&=&\left(\sum_{h\in \tilde{h}_\la^{(a)}}h\right)\left(\Op_{a,i_{Q-1}}-\Op_{i_1,i_{Q-1}}\right)\nonumber\\&=&(Q-2)!\sum_{\substack{i=1\\ i\neq a}}^Q\Op_{a,i}-(Q-3)!\sum_{\substack{i,j=1\\ i\neq j\neq a}}^{Q}\Op_{i,j}\nonumber \\&=&(Q-1)(Q-3)!\sum_{\substack{i=1\\ i\neq a}}^Q\Op_{a,i}+(Q-3)!\sum_{\substack{i=1\\ i\neq a}}^Q\Op_{i,a}-(Q-3)!\sum_{\substack{i,j=1\\i\neq j}}^{Q}\Op_{i,j} \,. \nonumber
\eea
We define the corresponding tensor $t_{a}^{[Q-1,1],1}$:
\bea t_{a}^{[Q-1,1],1}&\equiv& \tfrac{1}{(Q-1)(Q-3)!}e_{\la_Q}^{(a)}\Op_{a,i_{Q-1}} \label{taQ1-1-1} \\
\nonumber &=& \sum_{\substack{i=1\\ i\neq a}}^Q\Op_{a,i}+\tfrac{1}{Q-1}\sum_{\substack{i=1\\ i\neq a}}^Q\Op_{i,a}-\tfrac{1}{Q-1}\sum_{\substack{i,j=1\\i\neq j}}^{Q}\Op_{i,j} \,. \eea
However, we could also consider the following non-equivalent operation and define $t_{a}^{[Q-1,1],2}$:
\bea t_{a}^{[Q-1,1],2}&\equiv& \tfrac{1}{(Q-1)(Q-3)!}e_{\la_Q}^{(a)}\Op_{i_{Q-1},a} \label{taQ1-1-2} \\
\nonumber &=& \sum_{\substack{i=1\\ i\neq a}}^Q\Op_{i,a}+\tfrac{1}{Q-1}\sum_{\substack{i=1\\ i\neq a}}^Q\Op_{a,i}-\tfrac{1}{Q-1}\sum_{\substack{i,j=1\\i\neq j}}^{Q}\Op_{i,j} \,. \eea

A problematic feature in the definitions (\ref{taQ1-1-2})--(\ref{taQ1-1-2}) is that there is no manifest symmetry upon exchanging the two spins, $\sg_1$ and $\sg_2$. It thus appears natural to classify the operators acting also in terms of the group $\Sc_N$ permuting the $N$ spins. In the present case, with $N=2$, one possibility is to choose the symmetric representation $[2]$ of $\Sc_2$ and define the operator $t_a^{[Q-1,1],[2]}$:
\bea t_a^{[Q-1,1],[2]}&\equiv& \tfrac{Q-1}{Q}\left(t_{a}^{[Q-1,1],1}+t_{a}^{[Q-1,1],2}\right) \\
\nonumber &=& \sum_{\substack{i=1\\i\neq a}}^{Q}(\Op_{i,a}+\Op_{a,i})-\tfrac{2}{Q}\sum_{\substack{i,j=1\\i\neq j}}^{Q}\Op_{i,j}\eea
When expressed explicitly in terms of the spins, this operator reads
\be t_a^{[Q-1,1],[2]}(\sg_1,\sg_2)= \left. \begin{cases} \dt_{\sg_1,a}+\dt_{\sg_2,a}-\tfrac{2}{Q} & \text{for } \sg_1 \neq \sg_2 \,, \\ 0 & \text{for } \sg_1 = \sg_2 \,. \end{cases} \right. \ee
This precisely coincides with \eqref{magnetization2}, which was obtained in \cite{VJS2012,VJ2014} by imposing the symmetry between $\sg_1$ and $\sg_2$.

The other possibility is to choose the representation $[1,1]$ of $\Sc_2$ in which the two spins, $\sg_1$ and $\sg_2$, are antisymmetric. We therefore define the corresponding normalised tensor $t_a^{[Q-1,1],[1,1]}$:
\bea t_a^{[Q-1,1],[1,1]}&\equiv& \tfrac{Q-1}{Q-2}\left(t_{a}^{[Q-1,1],1}-t_{a}^{[Q-1,1],2}\right)\\
\nonumber &=& \sum_{\substack{i=1\\i\neq a}}^{Q}(\Op_{i,a}-\Op_{a,i}) \,. \eea
In terms of the spins this operator reads simply:
\be t_a^{[Q-1,1],[1,1]}(\sg_1,\sg_2)= \left. \begin{cases} \dt_{\sg_1,a}-\dt_{\sg_2,a} & \text{for } \sg_1 \neq \sg_2 \,, \\ 0 & \text{for } \sg_1 = \sg_2 \,. \end{cases} \right. \label{TQ1-1=1-1} \ee
This is our first example of a non-scalar operator that was not considered in the previous work \cite{VJ2014}.

It is straightforward to verify that the subsets $\left\{ t_a^{[Q-1,1],[2]} \, : a=1,\ldots,Q\right\}$ and $\left\{ t_a^{[Q-1,1],[1,1]} \, : a=1,\ldots,Q\right\}$ are indeed stable under the action of $\Sc_Q$, and of dimension $Q-1$.
We can rewrite the definition of both operators using an irreducible symmetriser of $\Sc_N$ (here with $N=2$) as
\bea t_a^{[Q-1,1],\la_N}=\tfrac{1}{\mathcal{N}}e_{\la_Q}^{(a)}\sn{e}_{\la_N}^{(a)}\Op_{a,i_{Q-1}} \,, \label{symm_bothN=2} \eea
where $\mathcal{N}$ is a normalisation constant. The operator $\sn{e}_{\la_N}^{(a)}$ acts according to \eqref{snaction} and is the irreducible symmetriser of the following Young tableau (depending on the two possible choices of $\la_N$):
\be
\ytableausetup{boxsize=2em,centertableaux}
\begin{ytableau}
1 & 2
\end{ytableau}
\quad \mbox{or} \quad
\ytableausetup{boxsize=2em,centertableaux}
\begin{ytableau}
1\\
 2 
\end{ytableau}
\ee

\subsubsection{Representation $[Q-2,2]$.}
We now construct the tensors with $2$ symmetric indices acting on $2$ spins. This corresponds to the Young diagram $\la_Q=[Q-2,2]$. Let us consider the following tableau
\be
\ytableausetup{boxsize=2em,centertableaux}
\begin{ytableau}
 i_1 & i_2 & \ldots & \, i_{Q-2} \, \\
 a_1 & a_2
\end{ytableau} \label{Q2-2-tableau}
\ee
where we again denote the permutation associated with this configuration by $a$.
The irreducible symmetriser is 
\be e_{[Q-2,2]}^{(a)}=\left(\sum_{h\in \tilde{h}_\la^{(a)}}h\right)\left(1+(a_1,a_2)\right)\left(1-(i_1,a_1)\right)\left(1-(i_2,a_2)\right) \,, \ee
where $\tilde{h}_\la^{(a)}$ now denotes the subset of $\Sc_Q$ consisting of all the permutations that leave $a_1$ and $a_2$ invariant. To generate the tensor we act with the irreducible symmetriser on $\Op_{a_1,a_2}$:
\begin{footnotesize}
\bea e_{[Q-2,2]}^{(a)} \Op_{a_1,a_2}&=& \left(\sum_{h\in h_\la^{(a)}}h\right)\left(\Op_{a_1,a_2}+\Op_{a_2,a_1}-\Op_{a_1,i_2}-\Op_{a_2,i_2}-\Op_{i_1,a_2}-\Op_{i_1,a_1}+2\Op_{i_1,i_2}\right)\nonumber
\\&=&(Q-2)!\left(\Op_{a_1,a_2}+\Op_{a_2,a_1}\right)-(Q-3)!\sum_{\substack{i=1\\ i\neq a_1,a_2}}^{Q}\left(\Op_{a_1,i}+\Op_{a_2,i}+\Op_{i,a_2}+\Op_{i,a_1}\right)\nonumber\\
&&+2(Q-4)!\sum_{\substack{i,j=1\\i\neq j\\i,j\neq a_1,a_2}}^{Q}\Op_{i,j} \,. \nonumber \eea
\end{footnotesize}
To get a nice expression we need to complete each sum with the missing terms, leading us to different prefactors. We normalise the resulting operator and define the corresponding tensor $t^{[Q-2,2]}_{a_1,a_2}$. The end result is:
\bea t^{[Q-2,2]}_{a_1,a_2}&=&\Op_{a_1,a_2}+\Op_{a_2,a_1}-\tfrac{1}{Q-2}\left(\sum_{\substack{i=1\\i\neq a_1}}^{Q}(\Op_{a_1,i}+\Op_{i,a_1})+\sum_{\substack{i=1\\i\neq a_2}}^{Q}(\Op_{a_2,i}+\Op_{i,a_2})\right)\nonumber\\&&+\tfrac{2}{(Q-1)(Q-2)}\sum_{\substack{i,j=1\\i\neq j}}^{Q}\Op_{i,j}\label{tensor2sym} \,. \eea
Note that despite of (\ref{Q2-2-tableau}), this expression can be extended to any values $a_1,a_2$, provided that $a_1\neq a_2$, by permuting the indices as required. The explicit expression of this tensor, omitting an overall factor of $(1-\delta_{\sg_1,\sg_2})$, is
\bea t^{[Q-2,2]}_{a_1,a_2}(\sg_1,\sg_2) &=& \dt_{a_1,\sg_1}\dt_{a_2,\sg_2}+\dt_{a_2,\sg_1}\dt_{a_1,\sg_2} \nonumber \\
 & & -\tfrac{1}{Q-2}\left(\dt_{a_1,\sg_1}+\dt_{a_1,\sg_2}+\dt_{a_2,\sg_1}+\dt_{a_2,\sg_2}\right)+\tfrac{2}{(Q-1)(Q-2)} \,. \label{symmetricN2}\eea
This coincides with \eqref{2cluster2}, as first found in \cite{VJS2012,VJ2014}. It is readily checked that the subspace $\left \{t^{[Q-2,2]}_{a_1,a_2} \, : \, 1\leq a_1\neq a_2\leq Q \right \}$ has the correct dimension $Q(Q-3)/2$, as given by the hook formula (\ref{hook-formula}) applied to the Young tableau $[Q-2,2]$.

It should be stressed that to obtain \eqref{symmetricN2} we did not specify any representation of $S_2$ for the spins $\sg_1$ and $\sg_2$. Indeed, for this tensor the chosen representations of $\Sc_Q$ and $\Sc_N$ are not independent: we cannot find operators with symmetric indices $a_1,a_2$ which are not symmetric for the spins $\sg_1,\sg_2$ as well. In fact, the symmetry of the spins $\sg$ is partially or totally dictated by the Young diagram $\lambda_Q$ with the first line removed. The general result about the relation between the $\Sc_Q$ and $\Sc_N$ symmetries will be stated below.

\subsubsection{Operators with anti-symmetric indices.}
To complete the discussion of operators acting on $N=2$ spins, we finally consider the tensors with two anti-symmetric indices. They correspond to the irreducible representation $[Q-2,1,1]$. Let us consider the following Young tableau
\be
\ytableausetup{boxsize=2em,centertableaux}
\begin{ytableau}
 i_1 & i_2 & \ldots & i_{Q-2} \\
 a_1 \\
 a_2
\end{ytableau}
\ee
with its corresponding irreducible symmetriser:
\be e_{[Q-2,1,1]}^{(a)}=\left(\sum_{h\in \tilde{h}_\la^{(a)}}h\right)\left(1-(i_1,a_1)-(i_1,a_2)-(a_1,a_2)+(i_1,a_1,a_2)+(i_1,a_2,a_1)\right) \,, \ee
where $\tilde{h}_\la^{(a)}$ denotes the subset of $\Sc_Q$ consisting of all  permutations that leave $a_1$ and $a_2$ invariant.
Acting on $\Op_{a_1,a_2}$ yields
\begin{footnotesize}
\bea e_{[Q-2,1,1]}^{(a)}\Op_{a_1,a_2} &=& (Q-2)!\left(\Op_{a_1,a_2}-\Op_{a_2,a_1}\right)-(Q-3)!\!\!\sum_{\substack{i=1\\i\neq a_1,a_2}}^Q\left(\Op_{a_1,i}\!-\!\Op_{i,a_1}\!+\!\Op_{i,a_2}\!-\!\Op_{a_2,i}\right) \nonumber \\
\nonumber &=&Q(Q-3)!\left(\Op_{a_1,a_2}-\Op_{a_2,a_1}\right) \\
\nonumber & & -(Q-3)!\sum_{\substack{i=1\\i\neq a_1}}^Q\left(\Op_{a_1,i}\!-\!\Op_{i,a_1}\right)-(Q-3)!\sum_{\substack{i=1\\i\neq a_2}}^Q\left(\Op_{i,a_2}\!-\!\Op_{a_2,i}\right) \,. \eea
\end{footnotesize}
We define the corresponding normalised operator $t_{a_1,a_2}^{[Q-2,1,1]}$:
\bea t_{a_1,a_2}^{[Q-2,1,1]}&=&\tfrac{1}{Q(Q-3)!}e_{[Q-2,1,1]}^{(a)}\Op_{a_1,a_2} \label{tensor2antisym} \\
&=& \Op_{a_1,a_2}-\Op_{a_2,a_1}-\tfrac{1}{Q}\left(\sum_{\substack{i=1\\i\neq a_1}}^Q\left(\Op_{a_1,i}\!-\!\Op_{i,a_1}\right)+\sum_{\substack{i=1\\i\neq a_2}}^Q\left(\Op_{i,a_2}\!-\!\Op_{a_2,i}\right)\right) \,. \nonumber \eea
Omitting again the factor $(1-\delta_{\sg_1,\sg_2})$, the explicit expression corresponding to \eqref{tensor2antisym} is
\be t_{a_1,a_2}^{[Q-2,1,1]}(\sg_1,\sg_2)=\dt_{\sg_1,a_1}\dt_{\sg_2,a_2}-\dt_{\sg_1,a_2}\dt_{\sg_2,a_1}-\tfrac{1}{Q}\left(\dt_{\sg_1,a_1}-\dt_{\sg_1,a_2}+\dt_{\sg_2,a_2}-\dt_{\sg_2,a_1}\right) \,.  \label{Q2-1-1=1-1} \ee

Also in this case the representation of $\Sc_Q$ here fully enforces the symmetry between the spins. This is a consequence of the fact that the number of boxes under the first line of $[Q-2,1,1]$ is $2$ and equal to the number of spins.

\subsubsection{Decomposition of $L_Q^{(2)}$.}
In the end we have the following decomposition of operators acting on $2$ spins:
\be L_Q^{(2)}= \underbrace{[Q]\oplus[Q-1,1]}_{L_Q^{(1)}}\oplus[Q]\oplus\underbrace{[Q-1,1]}_{[2]}\oplus\underbrace{[Q-1,1]}_{[1,1]}\oplus[Q-2,2]\oplus[Q-2,1,1] \,. \label{dimcheck2spins} \ee
The total dimension is seen to be $Q^2$, as it should. Two new operators were constructed, namely $t_{a_1,a_2}^{[Q-2,1,1]}$ and $t_{a}^{[Q_1,1],[1,1]}$. These operators were not discussed in \cite{VJS2012,VJ2014}, and omitting their contribution
to \eqref{dimcheck2spins} would give the dimension $Q(Q+1)/2$, corresponding to the number of generators of symmetric $Q \times Q$ matrices. Similarly, the contribution
of the two new operators to the dimension is $Q(Q-1)/2$, that is, the number of generators of anti-symmetric $Q \times Q$ matrices.

\subsection{Procedure for general representations}

The corresponding treatment of operators acting on $N=3$ spins can be found in \ref{sec:app3spins}, together with a general result on the dimension of the
space $L_Q^{(N)}$. From this, and the remarks given above, the procedure to build a tensor in an arbitrary representation of the group $\Sc_Q$ becomes clear.

The case of operators acting on $2$ spins in the representation $[Q-1,1]$ highlighted the need of taking into account the representation of the group $\Sc_N$ that dictates the symmetries of the $N$ spins. We need to be careful when choosing a representation of $\Sc_N$ because the representation $\la_Q$ already imposes symmetry constraints. We recall in particular the case $[Q-2,2]$ involving $2$ spins, where the constructed tensor \eqref{symmetricN2} automatically came out as being symmetric in $\sg_1$ and $\sg_2$, whereas the tensor of representation $[Q-1,1]$ acting on $2$ spins did not impose any $\Sc_2$ symmetry, which therefore needed to be subsequently imposed.

Let us consider now a Young diagram $\la_Q$ containing $Q-n$ boxes in the first row and a total of $n$ boxes in the remaining rows. From this diagram we are going to define a tensor with $n$ indices $a_1,\ldots,a_n$. We would like to define an operator acting on $N$ spins. This is obviously only possible if $N \ge n$, since we need a sufficient number of spins to act upon.

\subsubsection{Primal operator.}

If the condition $n=N$ is satisfied, the symmetry of the spins $\sg_i$ is entirely dictated by the Young diagram of shape $\la_Q$ with the first row removed. In this case we do not need to specify any $\la_N$ representation. We shall call such an operator \emph{primal}. We consider the Young tableau of shape $\la_Q$ where we insert all of the indices $a_1,\ldots,a_N$ in the boxes under the first row of $\la_Q$. This define a Young operator (irreducible symmetriser) via \eqref{irreduc-symm}. We impose that $a_1\neq a_2\neq\ldots\neq a_N$ in order to avoid redundancy with other representations, but we can relax this constraint in principle.

We now define the tensor $t^{\la_Q}_{a_1,\ldots,a_N}$ such that
\bea\label{definitionTensorPrimal} t^{\la_Q}_{a_1,\ldots,a_N}&=&\frac{1}{\Nc}e_{\la_Q}^{(a)}\Op_{a_1,\dots,a_N} \,, \eea
where $\Nc$ is an overall normalisation factor. Of course the normalisation is not fixed by representation theory and must be conveniently chosen (see later).

\subsubsection{Secondary operator.}
If $N > n$, we need to specify as well the Young diagram $\la_N$ that fixes the $\Sc_N$ representation. The corresponding operator will be referred to as \emph{secondary}. In order to obtain a non-zero tensor, $\la_N$ must obey a rule of consistency. Let us denote by $\widetilde{\la_Q}$ the Young diagram built from $\la_Q$ by removing its first row. We need to have the inclusion $\widetilde{\la_Q}\subseteq \la_N$, or, in other words, $\widetilde{\la_Q}$ must be obtained by removing some boxes of $\la_N$. We fix the representation of $\Sc_N$ by acting with the corresponding projector $\sn{e}_{\la_N}^{(a)}$, as in \eqref{symm_bothN=2}. In order to be coherent with the $\Sc_Q$ representation we need to change the definition of the latter projector with respect to \eqref{irreduc-symm}. We take, only for the $\Sc_N$ representation, the following definition: $\sn{e}_{\la_N}^{(a)}=a_{\la_N}^{(a)} s_{\la_N}^{(a)}$, where we note that the order of symmetrisation and anti-symmetrisation is now the opposite of that used in \eqref{irreduc-symm} (whence the tilde). The operator $t^{\la_Q,\la_N}_{a_1,\ldots,a_n}$ is then defined by
\bea\label{definitionTensor} t^{\la_Q,\la_N}_{a_1,\ldots,a_n}&=&\frac{1}{\Nc}e_{\la_Q}^{(a)}\sn{e}_{\la_N}^{(a)}\Op_{a_1,\dots,a_n,b_1,\ldots,b_{N-n}} \,, \eea
where $\Nc$ is an overall normalisation factor and $b_1,\ldots,b_{N-n}$ is a set of  unspecified indices. Once again, the normalisation is not fixed by representation theory, but is conveniently chosen so that the operator stays finite when $Q$ goes to infinity. We impose that $a_1\neq a_2\neq\ldots\neq a_n\neq b_1\neq\ldots\neq b_{N-n}$ in order to avoid redundancy with other representations, but again we can relax this constraint in principle.

We still need to specify the Young tableaux corresponding to the shapes $\la_Q$ and $\la_N$. First for $\la_Q$, we order $a_1,\ldots,a_N$ in the boxes strictly below the first row and place all the remaining values (including the $b$'s) in the first row. The exact values of the $b$'s are not important, since we symmetrise the indices which do not appear in the set $\{a_1,\ldots,a_n\}$.

Let us give an example with $\la_Q=[Q-3,2,1]$. The tensors have $3$ indices, $a_1,a_2$ and $a_3$, and the Young tableau has the form
\be
\ytableausetup{boxsize=2em,centertableaux}
\begin{ytableau}
 \ & \ &\  &\ldots & \ \\
 a_1 & a_2\\
 a_3 
\end{ytableau}
\ee
The tableau of shape $\la_N$ must respect the inner symmetry between $\sg_1,\ldots,\sg_n$ induced by $\la_Q$. As previously discussed, the $n$ first spins (among the $N$ defining the tensor) have already acquired a given symmetry, due to the action of $\la_Q$. To construct the tableau $\la_N$, we thus place the $n$ first indices as if we were considering the previous tableau of $\Sc_Q$ with its first row removed, and we choose an arbitrary index for each of the remaining $N-n$ boxes. For instance, if we want to act on $N=5$ spins with symmetry $[3,2]$ of $\Sc_N$ with a tensor belonging to the $\Sc_Q$ representation $[Q-3,2,1]$, we would have to choose a tableau of the following type
\be
\ytableausetup{boxsize=2em,centertableaux}
\begin{ytableau}
 1 & 2 &\ \\
 3 &\  
\end{ytableau}
\ee
and put arbitrarily the indices $4$ and $5$ in the blank boxes.

\subsubsection{Consistency between $\Sc_Q$ and $\Sc_N$ representations.}
\label{sec:consistency}

Even though the symmetries $\Sc_N$ and $\Sc_Q$ are different, they interact with each other and the two Young diagrams that we use to define a tensor must somehow be \emph{compatible}. Consider a tensor acting on $N$ spins with a given symmetry $\Sc_N$ associated to a Young tableau $\la_N$. We take $\la_Q$ to be a Young diagram with $Q$ boxes and define $\widetilde{\la}_Q$ as the diagram obtained by removing the boxes in the first line of $\la_Q$. A representation of $\Sc_Q$ leads to a non-trivial result if $\widetilde{\la}_Q$ can be obtained from $\la_N$ by \emph{removing at most one box in each column}. 

For instance consider the Young diagram $\lambda_N = [2,1]$. The only possible representations $\lambda_Q \in \Sc_Q$ that are compatible with $\lambda_N$ are $[Q-3,2,1]$, $[Q-2,2]$, $[Q-2,1,1]$ and $[Q-1,1]$. There are no tensors in the representation $[Q]$ since to obtain the corresponding $\widetilde{\lambda}_Q = \emptyset$ from $\lambda_N$ we would have to remove $2$ boxes from its the first column.

This consistency criterion can be understood from the following observation. Consider a pair of representations $\la_N$ and $\la_Q$ where $\la_Q$ is not compatible with $\la_N$ according to the previous rule. We construct a tensor $t^{\la_N,\la_Q}$ from the following definition:
\be t^{\la_Q,\la_N}=e_{\la_Q}\sn{e}_{\la_N}\Op_{a_1,a_2,\ldots,a_N} \,. \label{def_t_eQ_eN}
\ee
In general, $\sn{e}$ is given by its definition in terms of the action of permutations of $\Sc_N$ but if we act first with $\sn{e}$ we can write this operator in terms of permutations of $\Sc_Q$. We can interpret $\sn{e}_{\la_N}$ a sort of Young operator with a tableau of $N$ boxes filled with the indices $a_1,\ldots,a_N$ instead of the indices $1,\ldots,N$. 

For the sake of illustration, we consider $\la_N=[2,2]$ and $\la_Q=[Q-2,1,1]$. The tensor reads
\bea t^{\la_Q,\la_N}=e_{\la_Q}\sn{e}_{\la_N}\Op_{a_1,a_2,i_{Q-3},i_{Q-4}}\eea
and we use the following tableaux
\be
\ytableausetup{boxsize=2em,centertableaux}
\begin{ytableau}
 i_1 & i_2 & \ldots & \, i_{Q-2} \, \\
 a_1 \\
 a_2 
\end{ytableau}
\qquad 
\ytableausetup{boxsize=2em,centertableaux}
\begin{ytableau}
 1 & 3\\
 2 & 4
\end{ytableau}
\ee
for $\Sc_Q$ and $\Sc_N$ respectively. We can write $\sn{e}$ as permutations of $\Sc_Q$ with the operator associated to 
\be
\ytableausetup{boxsize=2em,centertableaux}
\begin{ytableau}
 a_1 & i_{Q-3}\\
 a_2 & i_{Q-2}
\end{ytableau}
\ee
Of couse this tableau is not a Young tableaux corresponding to a representation of $\Sc_Q$, but it defines an operator with permutations in a subgroup of $\Sc_Q$ equivalent to $\Sc_4$.

Since $\la_Q$ is not compatible with $\la_N$, there are $2$ indices $a_i$ and $a_j$ that are anti-symmetrised by $\sn{e}$ but belong to the first line of the tableau defining $e_{\la_Q}$ (here $i_{Q-3}$ and $i_{Q-2}$). We consider $Q$ large enough, so that those indices are pushed by definition somewhere in the first row of the tableau of shape $\la_Q$. This means that they
are first antisymmetrised by $\sn{e}_{\la_N}$ and then symmetrised by $e_{\la_Q}$. The result of these operations is exactly the null tensor, and hence justifies the consistency rule.

It is obvious that the argument carries over to the general case. If $\lambda_N$ contains a column having at least two more boxes than $\widetilde{\lambda_Q}$, then the corresponding indices will be antisymmetrised by $\sn{e}_{\la_N}$ and symmetrised by $e_{\la_Q}$, so that $t^{\la_Q,\la_N} = 0$ in (\ref{def_t_eQ_eN}).

\subsection{Internal structure and LCFT}

The structure of operators acting symmetrically on all their spins was initiated in \cite{VJS2012} and fully discussed in \cite{VJ2014}. The logarithmic features of the LCFT that arises in the continuum limit were studied by analysing how the divergences in the definition of the operators could be removed by mixing operators of the same scaling limit into Jordan cells.

For instance, the $2$-cluster operator with Young tableau $[Q-2,2]$ acting on $N=2$ spins is ill-defined for particular values of $Q$, as witnessed by the poles on the right-hand side of  \eqref{symmetricN2}. The mechanism explaining the logarithmic nature of correlation functions cures those divergences by mixing two operators \cite{VJS2012,VJ2014}. For instance, \eqref{symmetricN2} can be put into the form
\bea t^{[Q-2,2]}_{a,b} &=& \Op_{a_1,a_2}+\Op_{a_2,a_1} - \tfrac{1}{Q-2}\left(t^{[Q-1,1],[2]}_a+t^{[Q-1,1],[2]}_b\right)-\tfrac{2}{Q(Q-1)}t^{[Q]} \,, \label{sub2}\eea
in which the operator is obtained from the basis elements \eqref{basis-elements} by subtracting off components that belong to tensors with another symmetry. The equation \eqref{sub2} is ill-defined for percolation ($Q=1$) and this observation leads to the prediction that $t^{[Q-2,2]}_{a,b}$ is mixed with the energy operator $t^{[Q]}$ \eqref{invariant2}  in a Jordan cell of the dilatation operator for the CFT describing percolation. 

In order to make further progress on the LCFT structure of the Potts model we need to decompose all operators in a similar way. Consider a primal operator corresponding to a Young diagram $\la_Q$. All the tensors that are subtracted are called in the following subtensors and obey a simple rule. To be a subtensor (associated with a primal), its Young diagram $\la'_Q$ must be obtained by removing boxes of $\la_Q$ from the rows $2,3,\ldots$ and adding them to the first row. In particular, the subtensors are secondary to the primal from which they are subtracted. For example, in the case of $\la_Q=[Q-3,2,1]$ all the Young diagram verifying this condition are $[Q-2,2]$, $[Q-2,1,1]$
, $[Q-1,1]$ and $[Q]$. Moreover, if a subtensor corresponds to a diagram where {\em two or more} boxes were removed in the {\em same} column of $\la_Q$ it would not appear in the structure of the primal. In the example of $\la_Q=[Q-3,2,1]$, there is thus no subtensor with the symmetry of $\la'_Q=[Q]$ because in order to go from $\la_Q$ to $\la'_Q$ two boxes must be removed from the first column. This requirement can be understood in term of anti-symmetrisation/symmetrisation in the associated Young operator, using an argument similar to the one given in section~\ref{sec:consistency}.

The subtensors have the same $\lambda_N$ symmetry as the primal from which they are subtracted. Recall that we do not always need to express the $\lambda_N$ symmetry explicitly; see (\ref{sub2}) for an example. In particular, for a primal operator, $\lambda_N$ is simply obtained by removing the first row of $\lambda_Q$.

The internal structure that generalises \eqref{sub2} can thus be written in the symbolic form 
\bea t^{\la_Q}=(\Op)-\sum_{\la'_Q}\frac{1}{A_{\la_Q,\la'_Q}(Q)}t^{\la'_Q} \eea
where the sum is over all the Young diagram verifying the subtensor condition given above, $A_{\la_Q,\la'_Q}(Q)$ is a ratio of polynomials in $Q$, and $(\Op)$ is a combination of basis elements \eqref{basis-elements} that does not depend on $Q$ and whose exact form dictated by the Young tableau $\lambda_Q$ of the primal operator.

The exact form of $(\Op)$ and $t^{\la'_Q}$ can be computed in term of Young operators, but it is actually not necessary to do those explicit computations in order to make LCFT predictions. Rather, only the zeros of the functions $A_{\la_Q,\la'_Q}(Q)$ are needed in order to understand the mixing between operators.

For two Young diagrams $\la^1=[\la^1_0,\la^1_1,\ldots,\la^1_N]$ and $\la^2=[\la^2_0,\la^2_1,\ldots,\la^2_N]$, where $\la^i_j$ denotes the number of boxes%
\footnote{We use the convention of padding with zeros. More precisely, if $\la^i$ has $n+1$ rows with $n < N$, we set $\la^i_j = 0$ for $j > n+1$.}
in the $(j+1)$-th row of $\la^i$, we conjecture that---up to a multiplicative constant---the polynomial function $A_{\la^1,\la^2}(Q)$ is 

\be A_{\la^1,\la^2}(Q)\propto\prod_{i=1}^N\frac{\left(Q-n+i-1-\la^2_i\right)!}{\left(Q-n+i-1-\la^1_i\right)!}\label{conjecture} \ee
The multiplicative constant depends on the normalisation convention that we use in our definition \eqref{definitionTensor}.

The amount of evidence supporting the conjecture (\ref{conjecture}) will be discussed in section \ref{sec:generic-case} below.

\section{Correlation functions}
\label{correlationfunctions}
We now compute the two-point functions of the operators that we constructed in section~\ref{Potts} from symmetry considerations. The dependence of the correlation functions on the tensorial indices is fully determined by the irreducible representations of the group $\Sc_Q$. All the results will be related to the probability that two groups of spins, each comprising $N$ spins, are connected together by FK~clusters in a particular way.

For the purpose of taking the continuum limit, and for the physical interpretation as a two-point function to make sense, we imagine that each group of spins is confined to a small neighbourhood on the lattice. It should nevertheless be stressed that this constraint is completely irrelevant for the applicability of the representation theory.
For simplicity, we consider only operators that vanish if two spins are equal, and so an FK~cluster can at most contain one spin from each group. 

Let us introduce the notation used all along this section. Consider a correlation function of two operators, each of which acts on a group of $N$ spins situated in a small neighbourhood, centered at two points in space, that we denote $r_1$ and $r_2$ respectively. The correlations will involve the geometrical probability that some pairs of spins---always one from each neighbourhood---are connected in an FK~cluster. In order to represent such a probability we draw two groups of $N$ dots, the first group representing the spins $\sg_1^{(r_1)},\ldots,\sg_N^{(r_1)}$ near $r_1$, and the other $\sg_1^{(r_2)},\ldots,\sg_N^{(r_2)}$ near $r_2$. We draw a line between one spin at $r_1$ and one spin at $r_2$ if they belong to the same FK~cluster. Of course two spins in the same neighbourhood cannot be connected, since we consider exclusively operators that vanish if anyone of the $N$ spins acted upon coincide.

For instance, let us illustrate this with an operator $t(\sg_1,\sg_2,\sg_3)$ acting on $N=3$ spins and its two-point correlation function $\left<t(r_1)t(r_2)\right>$. The symbol $\Pex$ denotes the probability of the following situation: the spin $\sg_1^{(r_1)}$ and the spin $\sg_2^{(r_2)}$ are not connected to the four others through an FK~cluster, whereas $\sg_2^{(r_1)}$ (resp.\ $\sg_3^{(r_1)}$) and $\sg_1^{(r_2)}$ (resp.\ $\sg_3^{(r_2)}$) are in the same FK~cluster. This geometrical dependence, of major importance for the physical system, will depend on all the information concerning the lattice, in particular the number of dimensions $d$ in which it lives. Unfortunately, computing such a quantity exactly for very general cases is out of reach,%
\footnote{Exact expressions exist for simple cases in two dimensions, such as for the Ising model. But even this is often restricted to specific regular lattices (such as the square lattice) with specific boundary conditions (free, cylindrical, toroidal, {\em etc}). However, in $d=2$, CFT can often predict the power-law asymptotic form of such correlation functions, and this turns out to be universal (i.e., independent of lattice details). The appearance of logarithms will be discussed below.}
but thanks to representation theory, and later scale invariance, we are still able to understand some features. 

In order to compute the correlation function between two operators, we list all possible connectivities and compute their respective amplitudes. The average is taken by summing the explicit expressions of the product of operators over independent groups of spins. In other words, if two spins are connected by an FK~cluster, we set them equal, and the average is taken by independently summing over the resulting set of FK~clusters.

Our results comprise the two-point correlation functions found in \cite{VJ2014}, except that we now extend the procedure to our new, non-scalar operators. We will find interesting properties that do not appear when one considers only symmetric operators. On the other hand, \cite{VJ2014} discussed also some cases of three-point correlation functions---although this can also be done in the present context, we have chosen to focus on two-point functions only.

Let us first investigate the simplest new case corresponding to an anti-symmetric tensor acting on 2 spins. This example will also serve to illustrate the way in which correlation functions are computed. We will then comment briefly on the case of tensors with mixed symmetry (i.e., neither fully symmetric, nor fully anti-symmetric).

\subsection{Symmetric operator acting on $2$ spins}
The two-point function of the $2$-cluster operator was computed in \cite{VJS2012,VJ2014}. We compute it again here in order to illustrate the method and emphasise a particular feature. Let us thus consider the average
\be
 \left<t^{[Q-2,2]}_{a_1,a_2}(\sg_1,\sg_2)t^{[Q-2,2]}_{b_1,b_2}(\sg_3,\sg_4)\right> \,.
\ee
We first compute the amplitude over the probability that $\sg_1$ and $\sg_3$ are in the same cluster, but that $\sg_2$ and $\sg_4$ are in distinct independent clusters. This probability is represented by $\Panan$. Its amplitude is 
\be \tfrac{1}{Q^3}\sum_{\sg_1,\sg_2,\sg_4}t^{[Q-2,2]}_{a_1,a_2}(\sg_1,\sg_2)t^{[Q-2,2]}_{b_1,b_2}(\sg_1,\sg_4) =0\ee
The computation is straightforward; one only needs to insert the expression \eqref{symmetricN2} and remember that $t^{[Q-2,2]}(\sg_1,\sg_2)=0$ when $\sg_1=\sg_2$.%
\footnote{Alternatively, one simply notices that $\sum_{\sigma_2} t^{[Q-2,2]}_{a_1,a_2}(\sg_1,\sg_2) = 0$.}
Computing in the same way the amplitude of each possible probability of having the 4 spins connected in any specific way, we see that only two amplitude are non-zero. They correspond to the probabilities $\Ps$ and $\Pc$. They have the same amplitude, which is 
\bea \tfrac{1}{Q^2}\sum_{\sg_1,\sg_2}t^{[Q-2,2]}_{a_1,a_2}(\sg_1,\sg_2)t^{[Q-2,2]}_{b_1,b_2}(\sg_1,\sg_2) = \nonumber \\
 \tfrac{2}{Q^2}\Big(\dt_{a_1,b_1}\dt_{a_2,b_2}+\dt_{a_1,b_2}\dt_{a_2,b_1} - \tfrac{1}{Q-2}\Big(\dt_{a_1,b_1}+\dt_{a_2,b_2}\nonumber+\dt_{a_1,b_2}+\dt_{a_2,b_1}\Big)+\tfrac{2}{(Q-2)(Q-1)}\Big) \,. \eea
In the end the correlation function is 
\bea
\label{2points[2]}\left<t^{[Q-2,2],[2]}_{a_1,a_2}(r_1)t^{[Q-2,2],[2]}_{b_1,b_2}(r_2)\right> = \tfrac{2}{Q^2}\Big(\dt_{a_1,b_1}\dt_{a_2,b_2}+\dt_{a_1,b_2}\dt_{a_2,b_1} \nonumber \\
-\tfrac{1}{Q-2}\Big(\dt_{a_1,b_1}+\dt_{a_2,b_2}+\dt_{a_1,b_2}+\dt_{a_2,b_1}\Big)+\tfrac{2}{(Q-2)(Q-1)}\Big)\left(\Ps+\Pc\right) \,, \eea
where  $r_1$ and $r_2$ are respectively the position of the neighbourhood of $(\sg_1,\sg_2)$ and $(\sg_3,\sg_4)$.

\subsection{Anti-symmetric operator acting on $2$ spins}
Next consider the correlation function for a tensor with $N=2$ spins in the symmetry corresponding to the representation of $\Sc_N$ given by the anti-symmetric Young tableau $[1,1]$. It can be computed using the same procedure as above. Note that in the decomposition of $t^{[Q-2,1,1],[1,1]}_{a,b}$ given by \eqref{Q2-1-1=1-1}
appears the subtensor $t^{[Q-1,1],[1,1]}_a$ given by \eqref{TQ1-1=1-1}; the latter corresponds to a tensor in the standard representation, but acting on two anti-symmetric spins.%
\footnote{In this section we have set $t^{[Q-2,1,1],[1,1]}_{a,b} \equiv t^{[Q-2,1,1]}_{a,b}$, adding a superfluous $[1,1]$ to the general notation for extra clarity.}
We also compute its correlation function for completeness. We find
\begin{footnotesize}
\bea\label{2points[1][1,1]}\left<t^{[Q-1,1],[1,1]}_{a}(r_1)t^{[Q-1,1],[1,1]}_{b}(r_2)\right> =   \\
 \tfrac{1}{Q}\Big(\dt_{a,b}-\tfrac{1}{Q}\Big)\Bigg(2\,\Ps-2\,\Pc+\left( \Panan+\Pnbnb -\Pnban-\Pannb\right)\Bigg) \nonumber \,,
\\ \label{2points[1,1]} \left<t^{[Q-2,1,1],[1,1]}_{a_1,a_2}(r_1)t^{[Q-2,1,1],[1,1]}_{b_1,b_2}(r_2)\right> =  \\
 \tfrac{8}{Q^2}\Big(\dt_{a_1,b_1}\dt_{a_2,b_2}-\dt_{a_1,b_2}\dt_{a_2,b_1}-\tfrac{1}{Q}\Big(\dt_{a_1,b_1}+\dt_{a_2,b_2} -\dt_{a_1,b_2}-\dt_{a_2,b_1}\Big)\Big)\left(\Ps-\Pc\right) \,, \nonumber
\\ \left<t^{[Q-2,1,1],[1,1]}_{a_1,a_2}(r_1)t^{[Q-1,1],[1,1]}_{b}(r_2)\right> = 0 \,. \label{2points[1][1,1]with[1,1]} \eea
\end{footnotesize}
Before commenting on this result, let us again explain carefully how we can compute these quantities. Consider the two-point function $\left<t^{[Q-2,1,1],[1,1]}_{a_1,a_2}(r_1)t^{[Q-2,1,1],[1,1]}_{b_1,b_2}(r_2)\right>$ of operators whose expression was computed previously in \eqref{Q2-1-1=1-1}:
\be t^{[Q-2,1,1],[1,1]}_{a_1,a_2}(\sg_1,\sg_2)=\dt_{a_1,\sg_1}\dt_{a_2,\sg_2}-\dt_{a_1,\sg_2}\dt_{a_2,\sg_1}-\tfrac{1}{Q}\left(\dt_{a_1,\sg_1}+\dt_{a_2,\sg_2}-\dt_{a_1,\sg_2}-\dt_{a_2,\sg_1}\right) \,. \nonumber \ee
When we compute the correlation function, three cases appear. First, supposing that the spins around $r_1$ are not connected to the spins around $r_2$, the two-point function is proportional to the probability $\Pnn$. The amplitude of this probability is obtained after averaging the product of the two tensors summed independently over all the spins:%
\footnote{Alternatively, one simply notices that $\sum_{\sigma_2} t^{[Q-2,1,1],[1,1]}_{a_1,a_2}(\sg_1,\sg_2) = 0$.}
\be \tfrac{1}{Q^2}\sum_{\sg_1,\sg_2}t^{[Q-2,1,1],[1,1]}_{a_1,a_2}(\sg_1,\sg_2)\ \tfrac{1}{Q^2}\sum_{\sg_3,\sg_4}t^{[Q-2,1,1],[1,1]}_{b_1,b_2}(\sg_3,\sg_4)=0 \,. \ee

The second case involves situations where a unique spin around $r_1$ is connected to another spin around $r_2$. There are four such situations, related to $\Panan$,~$\Pnbnb$,~$\Pannb$~and~$\Pnban$. Let us illustrate the first case. In order to compute the amplitude of this probability appearing in the correlation function, we need to consider the product $t^{[Q-2,1,1],[1,1]}_{a_1,a_2}(\sg_1,\sg_2)t^{[Q-2,1,1],[1,1]}_{b_1,b_2}(\sg_3,\sg_4)$, average independently over $\sg_2$ and $\sg_4$, and average over $\sg_1$ and $\sg_3$ with the constraint that they are in the same FK cluster (whence $\sg_1=\sg_3$). It follows that
\be \tfrac{1}{Q^3}\sum_{\sg_1,\sg_2,\sg_4}t^{[Q-2,1,1],[1,1]}_{a_1,a_2}(\sg_1,\sg_2)\ t^{[Q-2,1,1],[1,1]}_{b_1,b_2}(\sg_1,\sg_4)=0 \,. \ee
The corresponding amplitude in \eqref{2points[1,1]} is thus zero. However, if we consider the two-point correlation function involving two tensors $t^{[Q-2,1],[1,1]}_a$ as in \eqref{2points[1][1,1]}, we find a non-trivial result for that same amplitude:
\be \tfrac{1}{Q^3}\sum_{\sg_1,\sg_2,\sg_4}t^{[Q-1,1],[1,1]}_{a}(\sg_1,\sg_2)\ t^{[Q-1,1],[1,1]}_{b}(\sg_1,\sg_4)=\tfrac{1}{Q}\left(\dt_{a,b}-\tfrac{1}{Q}\right) \,. \ee

Finally, consider the case with two pairwise connections by FK~clusters: $\Ps$, $\Pc$. Averaging the tensor product over the spins with these constraints, we get
\bea \tfrac{1}{Q^2}\sum_{\sg_1,\sg_2}t^{[Q-2,1,1],[1,1]}_{a_1,a_2}(\sg_1,\sg_2)\ t^{[Q-2,1,1],[1,1]}_{b_1,b_2}(\sg_1,\sg_2)= \nonumber \\
 \tfrac{8}{Q^2}\Big(\dt_{a_1,b_1}\dt_{a_2,b_2}-\dt_{a_1,b_2}\dt_{a_2,b_1}- \tfrac{1}{Q}\Big(\dt_{a_1,b_1}+\dt_{a_2,b_2}-\dt_{a_1,b_2}-\dt_{a_2,b_1}\Big)\Big) \,, \eea
or the same expression with the opposite sign if we permute $\sg_1$ and $\sg_2$ in one of the two tensors. Performing a similar computation for the two-point function of
$t^{[Q-2,1],[1,1]}_a$ completes the proof of \eqref{2points[1][1,1]}--\eqref{2points[1,1]}. The crossed correlator \eqref{2points[1][1,1]with[1,1]} of the two different tensors with $[1,1]$ symmetry turns out to vanish identically.

We notice that an operator that corresponds to a representation of $\Sc_Q$ having $n$ tensorial indices, involves only probabilities that have at least $n$ FK~clusters. This result is generally true for any representation of $\Sc_Q$ and $\Sc_N$ (and was already noticed for the symmetric case in \cite{VJ2014}). In particular, when a tensor acts on the minimum number of spins (i.e., when $n=N$), such as in \eqref{2points[1,1]}, the result involves only probabilities with all the spins connected in a certain FK~configuration.
In other words, primal tensors of rank $N$ (i.e., with $N$ tensor indices) are precisely the $N$-cluster operators discussed in section~\ref{sec:Potts-FK}, having a symmetry given by the corresponding Young diagram $\lambda_Q \in \Sc_Q$.

The major difference between \eqref{2points[1,1]} and the correlation function \eqref{2points[2]} found for the symmetric tensor with $N=2$ \cite{VJS2012,VJ2014}, is that the dominant term is not the probability to get a cluster that expands from $0$ to $r$, but a sub-leading contribution to that probability. Indeed, the minus sign between the two FK~diagrams cancels out the dominant power law decay of the two-cluster operator, leaving only a much smaller correction term. A similar situation will appear in the next section when we consider a mixed symmetry.

\subsection{Operator with mixed symmetry : $[Q-3,2,1]$}
Briefly, and without giving the details of the computations, we present here the simplest case of mixed symmetry, namely the representation of $\Sc_Q$ $[Q-3,2,1]$ with $\la_N=[2,1]$. After computation we find
\bea \left<t^{[Q-3,2,1],[2,1]}_{a_1,a_2,a_3}(r_1)t^{[Q-3,2,1],[2,1]}_{b_1,b_2,b_3}(r_2)\right>= \nonumber \\
\tfrac{9}{Q^3}\Big(\dt^3 +\tfrac{1}{2(Q-1)}\dt^2_s-\tfrac{3}{2(Q-3)}\dt^2_a+\tfrac{1}{(Q-1)(Q-3)}\dt^1\Big) \Big) \label{Corr[2,1]} \\
\times\left(2\ \Pabc+\Pbac +\Pacb-2\ \Pcba-\Pbca-\Pcab\right) \nonumber \,,
\eea
where $\dt^n_i$ is a short-hand notation for a sum of products of $n$ Kronecker functions. The terms $\dt^3$, $\dt^2_s$, $\dt_a^2$ and $\dt^1$ involve respectively $6$, $18$, $8$ and $9$ terms.

Let us clarify the important features of this result. First, there is an overall (and rather unimportant) constant which is not fixed by symmetry considerations. This multiplies a factor containing an appropriate function of the tensor indices, and another factor which is an appropriate linear combination of FK probabilities. These two factors are entirely determined by the representation in which the tensor being considered lives. These factors are seen to be more complicated than in the fully symmetric case \eqref{2points[2]} or the anti-symmetric case \eqref{2points[1,1]}.
We notice that, as we have discussed before, since the considered tensor has an $\Sc_Q$ representation that is fully redundant with its $\Sc_N$ representation ($n=N$), we only get probabilities where all the spins at $r_1$ are connected through FK~clusters to another spin at $r_2$. In other words, there are no terms such as $\Pabn$ or $\Pcbn$, etc.

Again the mixed correlation functions of two operators, living in the same $\Sc_N$ representation but with a different $\Sc_Q$ symmetry, are exactly vanishing:
\begin{align} \left<t^{[Q-3,2,1],[2,1]}_{a_1,a_2,a_3}(r_1)t^{[Q-2,2],[2,1]}_{b_1,b_2}(r_2)\right> &= 0 \,, \nonumber\\
 \left<t^{[Q-3,2,1],[2,1]}_{a_1,a_2,a_3}(r_1)t^{[Q-2,1,1],[2,1]}_{b_1,b_2}(r_2)\right> &= 0 \,, \nonumber\\
 \left<t^{[Q-3,2,1],[2,1]}_{a_1,a_2,a_3}(r_1)t^{[Q-1,1],[2,1]}_{b}(r_2)\right> &= 0 \,. \nonumber\end{align}

We can also compute the correlation functions involving the two other operators corresponding to the $\Sc_Q$ representations $[Q-3,3]$ and $[Q-3,1,1,1]$. They again vanish:
\begin{align}
 \left<t^{[Q-3,2,1],[2,1]}_{a_1,a_2,a_3}(r_1)t^{[Q-3,3],[3]}_{b_1,b_2,b_3}(r_2)\right> &=0 \,, \nonumber\\
 \left<t^{[Q-3,2,1],[2,1]}_{a_1,a_2,a_3}(r_1)t^{[Q-3,1,1,1],[1,1,1]}_{b_1,b_2,b_3}(r_2)\right> &=0 \,.
\end{align}

Let us mention that in general the correlation function between two operators in the {\em same} $\Sc_Q$ representation but with a different $\la_N$ Young tableaux is non-vanishing. When we compute $\left<t^{[Q-2,2],[2,1]}_{a_1,a_2}(r_1)t^{[Q-2,2],[3]}_{b_1,b_2}\right>$ we find a result with probabilities involving $2$ propagating FK clusters. On the other hand, the vanishing of correlation functions between field in {\em different} $\Sc_Q$ representations is consistent with general representation theoretical expectations.

\subsection{Generic case}
\label{sec:generic-case}
Considering the fact that \eqref{Corr[2,1]} is the easiest two-point function of tensors with a mixed symmetry, it is clear that other cases cannot easily be computed by hand.
We have therefore used exact symbolic computations in {\sc Mathematica} extensively in order to construct the tensors in higher representations and their corresponding two-point functions.
In this way, we have computed all tensor acting on up to $N=5$ spins. We give in the appendix \ref{sec:app3spins} the expressions for the operators acting on $N=3$ spins and their two-point functions. We have extensive evidence for the conjecture \eqref{conjecture} since we are able to compute numerically the exact decomposition to obtain the poles. Numerical evidence and construction of the operators are available in electronic form as supplementary material to this paper.\footnote{In the form of a file {\tt TensorsPotts.nb} that can be processed by {\sc Mathematica}.}

For tensors corresponding to primal operators, the correlation function of two identical fields is generically found to be composed of two factors, apart from the overall normalisation. The first factor depends on the tensor indices and is entirely fixed by the chosen $\Sc_Q$ representation. The second factor is a linear combination of probabilities that $N$ distinct FK clusters propagate from one neighbourhood to the other, and where the order of the cluster end points depends also on the representation. This latter part depends on the specific lattice, and the dimension $d$. Computing the decay of this linear combination of probabilities determines the corresponding critical exponent, and is amenable to numerical work (see below). The required linear combination can be obtained for generic $\la_N$ very easily by acting with the irreducible symmetriser $\sn{e}_{\la_N}$ on cluster end points.

\section{Physical interpretation}

\subsection{Primal and secondary operators}

In $d=2$ dimensions, operators in conformal field theories can be classified as primaries, quasi-primaries and descendants, according to their covariance properties under (local) conformal transformations. In $d>2$, the conformal algebra is finite-dimensional and strictly speaking operators are at most quasi-primaries. The operators that we labelled \emph{primal} are believed to correspond to primaries in $d=2$ or quasi-primaries in $d>2$. This interpretation is supported by the vanishing two-point functions between two different primal operators. The operators labelled as secondary will in general correspond to descendants or sub-leading operators in $d>2$.

There appears to be at least one exception to this general correspondence. The energy operator corresponds \cite{VJ2014} to the secondary tensor discussed in section \ref{sec:energy-operator}, with $\la_Q=[Q]$ and $\la_N=[2]$, which is a well-known to be a primary operator in $d=2$ CFT.

\subsection{Critical exponents on a cylinder}

In this section we discuss the identification of the tensors constructed in section~\ref{Potts} with conformal fields in $d=2$ dimensions. The operators with $n=N$ that we have dubbed primal will be shown to correspond to primary fields in the sense of CFT. All the corresponding scaling dimensions will be identified exactly. We shall also see that the $d=2$ case conceals a subtlety with respect to the case of arbitrary dimension, $d > 2$. Namely, depending on the precise lattice regularisation used to define the FK~probabilities, that enter into the two-point correlators, it is in some cases possible to obtain different results for the critical exponents.

The key remark is that in $d=2$, some configurations of connectivities between clusters are not possible. Consider the usual situation with two small neighbourhoods, ${\cal D}_1$ and ${\cal D}_2$, each containing $N$ spins. We shall suppose each neighbourhood ${\cal D}$ to be a simply connected domain, and the $N$ marked points to reside at its boundary $\partial {\cal D}$. On a given lattice, if the $N$ marked points cover the boundary of ${\cal D}$ sufficiently tightly, the FK cluster emanating from some particular marked point $i \in \partial {\cal D}$  will be unable to enter the interior of ${\cal D}$ and exit it in-between two other marked points, $j$ and $k$. Thus, since clusters cannot cross each other in $d=2$, this implies that the $N$ marked FK clusters will only be able to permute cyclically in the space outside ${\cal D}$. In particular, the clusters cannot undergo arbitrary permutations.

To appreciate this remark, consider the specific example shown below, on the triangular lattice. We take each of ${\cal D}_1$ and ${\cal D}_2$ to consist of the $3$ vertices around an elementary triangle. Consider now a pair of $3$-cluster operators, ${\cal O}_1$ and ${\cal O}_2$:
\be
\psset{unit=1.5cm}
\begin{pspicture}[shift = 0](0,0)(6,3.5)
\multirput(0,0)(0,1.732){2}{
\multirput(0,0)(1,0){6}{\psline[linewidth = 1pt,linecolor = black](0,0)(1,0)\psline[linewidth = 1pt,linecolor = black](0,0)(0.5,0.866)\psline[linewidth = 1pt,linecolor = black](1,0)(0.5,0.866)}
}
\multirput(0,0)(0,1.732){2}{
\multirput(0,0.866)(1,0){6}{\psline[linewidth = 1pt,linecolor = black](0,0.866)(1,0.866)\psline[linewidth = 1pt,linecolor = black](0,0.866)(0.5,0)\psline[linewidth = 1pt,linecolor = black](1,0.866)(0.5,0)}
}
\multirput(0,0)(0,1.732){2}{
\psline[linewidth = 1pt,linecolor = black](0.5,0.866)(5.5,0.866)
}
\pscircle[fillstyle=solid, fillcolor=red](1.5,0.866){0.1}
\pscircle[fillstyle=solid, fillcolor=red](1,1.732){0.1}
\pscircle[fillstyle=solid, fillcolor=red](2,1.732){0.1}
\rput(1.5,1.4){\footnotesize$\Op_1$}
\rput(1.5,1.15){\footnotesize$1$}
\rput(1.25,1.6){\footnotesize$2$}
\rput(1.75,1.6){\footnotesize$3$}

\pscircle[fillstyle=solid, fillcolor=red](4.5,2.598){0.1}
\pscircle[fillstyle=solid, fillcolor=red](4,1.732){0.1}
\pscircle[fillstyle=solid, fillcolor=red](5,1.732){0.1}
\rput(4.5,2){\footnotesize$\Op_2$}
\rput(4.5,2.35){\footnotesize$1$}
\rput(4.25,1.9){\footnotesize$2$}
\rput(4.75,1.9){\footnotesize$3$}
\end{pspicture}
\ee
It is clear that in this setup the probabilities $\Pbac$, $\Pacb$ and $\Pcba$ are exactly $0$. This would not be the case in higher dimensions, since the clusters could cross freely to realise arbitrary permutations, by ``entering the third dimension''.

While this setup is relevant for numerical simulations of the Monte Carlo type, the same conclusion applies to transfer matrix diagonalisations. Indeed, in the latter, the two boundaries, $\partial {\cal D}_1$ and $\partial {\cal D}_2$ are pushed to opposite extremites of an infinite cylinder, by means of a conformal mapping. So also in this geometry, only cyclic permutations of the marked clusters can be realised, because the clusters propagate from one cylinder extremity to the other and can only interchange by winding around the periodic boundary condition.

In other words, the permutation group is not fully relevant in $d=2$ dimensions. Therefore, if we consider for instance the mixed Young diagram $[Q-3,2,1]$, the corresponding two-point function will only involve the following linear combination of probabilities:
 \bea \left<t^{[Q-3,2,1],[2,1]}_{a_1,a_2,a_3}(r_1)t^{[Q-3,2,1],[2,1]}_{b_1,b_2,b_3}(r_2)\right>\propto\left(2\ \Pabc-\Pbca-\Pcab\right)
\eea
as opposed to \eqref{Corr[2,1]} in the general case.

Following these observations, consider now an irreducible representation of $\Sc_N$. Within the subgroup ${\mathbb Z}_N$, this is reducible and can be decomposed into a direct sum of irreducible representations of ${\mathbb Z}_N$. The operators corresponding to each term in this decomposition are well known, so we will obtain in particular an identification of the critical exponents of our operators within $d=2$ CFT. We recall that irreducible representations of ${\mathbb Z}_N$ are all of dimension $1$ and labelled by an integer $p$, with $0\leq p\leq N-1$, where the first cyclic permutation is represented by $\exp \left( i \frac{2\pi p}{N} \right)$.

A procedure for decomposing an irreducible representation of $\Sc_N$ into representations of ${\mathbb Z}_N$ is given for a more general case in \cite{S89}. Given a Young diagram $\lambda_N \in \Sc_N$ the procedure in our case is the following:
\begin{itemize}
\item List all the $D$ standard Young tableaux corresponding to the Young diagram $\la_N$. Here, $D$ denotes the dimension of the irreducible representation corresponding to $\la_N$, which can be computed from \eqref{hook-formula}.
\item Compute the \emph{index} of each standard Young tableau. It is defined as the sum of the {\em descents} of the Young tableau. A number $i$ is a descent, if $i+1$ appears in a row strictly below $i$ in the tableau.
\item Let $p$ denote the index modulo $N$, so that it obeys $\frac{-N}{2} <p\leq \frac{N}{2}$.
\item Repeating this for all $D$ standard Young tableaux, we obtain $D$ such integers: $p_1,p_2,\ldots,p_D$. An irreducible representation of $\Sc_N$ restricted to the cyclic group ${\mathbb Z}_N$ can then be decomposed as follows:
\be \la_N= \exp \left( i \frac{2\pi p_1}{N} \right) \oplus \exp \left( i \frac{2\pi p_2}{N} \right) \oplus\ldots\oplus \exp \left( i \frac{2\pi p_D}{N} \right) \,. \ee
\end{itemize}
We call $p/N$ the pseudo-momentum. The critical exponents of the operator within this representation of ${\mathbb Z}_N$ are known \cite{ReadSaleur01,GRSV}.

They can be written in terms of an extended Kac table \cite{FQS} of conformal weights 
\be h_{r,s}=\frac{(r(x+1)-sx)^2-1}{4x(x+1)} \,, \ee
with Kac labels $(r,s)$ of the form $(r,s)=(\pm \tfrac{p}{N},N)$, as follows:
\bea \Delta_{p,N} &=& h_{\frac{p}{N},N}+h_{-\frac{p}{N},N} = \frac{N^2(N^2 x^2-1) + p^2(x+1)^2}{2 N^2 x(x+1)} \,, \label{Delta_pN} \\
\ell &=& h_{-\frac{p}{N},N}-h_{\frac{p}{N},N} = p \,. \eea
Here $\Delta_{p,N}$ denotes the scaling dimension, and $\ell$ is the conformal spin. The number of states in the Potts model corresponds to
the parameter $x$ via
\be
 Q = 4 \cos^2 \left( \frac{\pi}{x+1} \right) \,.
\ee

In the end, the operator corresponding to the irreducible representation $\la_N$ will have a dominant scaling dimension corresponding to the smallest exponent in the set $\{\Delta_{|p_i|,N}\}$, where the $p_i$ correspond to the decomposition of $\la_N$ restricted to the cyclic group ${\mathbb Z}_N$. Since $p\rightarrow\Delta_{p,N}$ is an increasing function for $p$ positive, we find that the leading contribution comes from the smallest absolute value of the $p$'s.

For instance, the representation of $\Sc_2$ with Young diagram $[1,1]$ is of dimension $1$ and has only one standard Young tableau:
\be
\ytableausetup{boxsize=2em,centertableaux}
\begin{ytableau}
 $1$  \\
 $2$ 
\end{ytableau}
\ee
We find that $p=1$, so the scaling dimension is $\Delta_{1,2}$ and its spin is $\ell = 1$.

Let us give a slightly more complicated example, from $\Sc_4$, with the Young diagram $[2,2]$. There are two standard tableaux:
\be
\ytableausetup{boxsize=2em,centertableaux}
\begin{ytableau}
 $1$  & $2$ \\
 $3$ & $4$
\end{ytableau}
\qquad
\begin{ytableau}
 $1$  & $3$ \\
 $2$  & $4$ 
\end{ytableau}
\ee
We find $p = 2$ for the first tableau and $p=0$ for the second. The dominant contribution to the two-point function of the operator associated with this representation, in $d=2$ CFT, will have scaling dimension  $\Delta_{0,4}$ and spin $\ell = 0$. The scaling dimension $\Delta_{2,4}$, corresponding to the first tableau, is a sub-leading exponent.
 
 The only exception to this rule is for the case $N=1$ where the magnetisation operator is given by $(r,s)=(\tfrac{1}{2},0)$ instead of $(r,s)=(0,1)$ \cite{VJ2014}.
 Indeed, the $N$-cluster operator corresponds generically (i.e., for $N > 1$) to a $2N$-leg watermelon operator in $d=2$, but this is not true for $N=1$.

\subsection{Numerics}

Numerical checks of the analytically predicted critical exponents can be performed by using several methods.

\begin{figure}[h!]
\centering
\includegraphics[width=14cm]{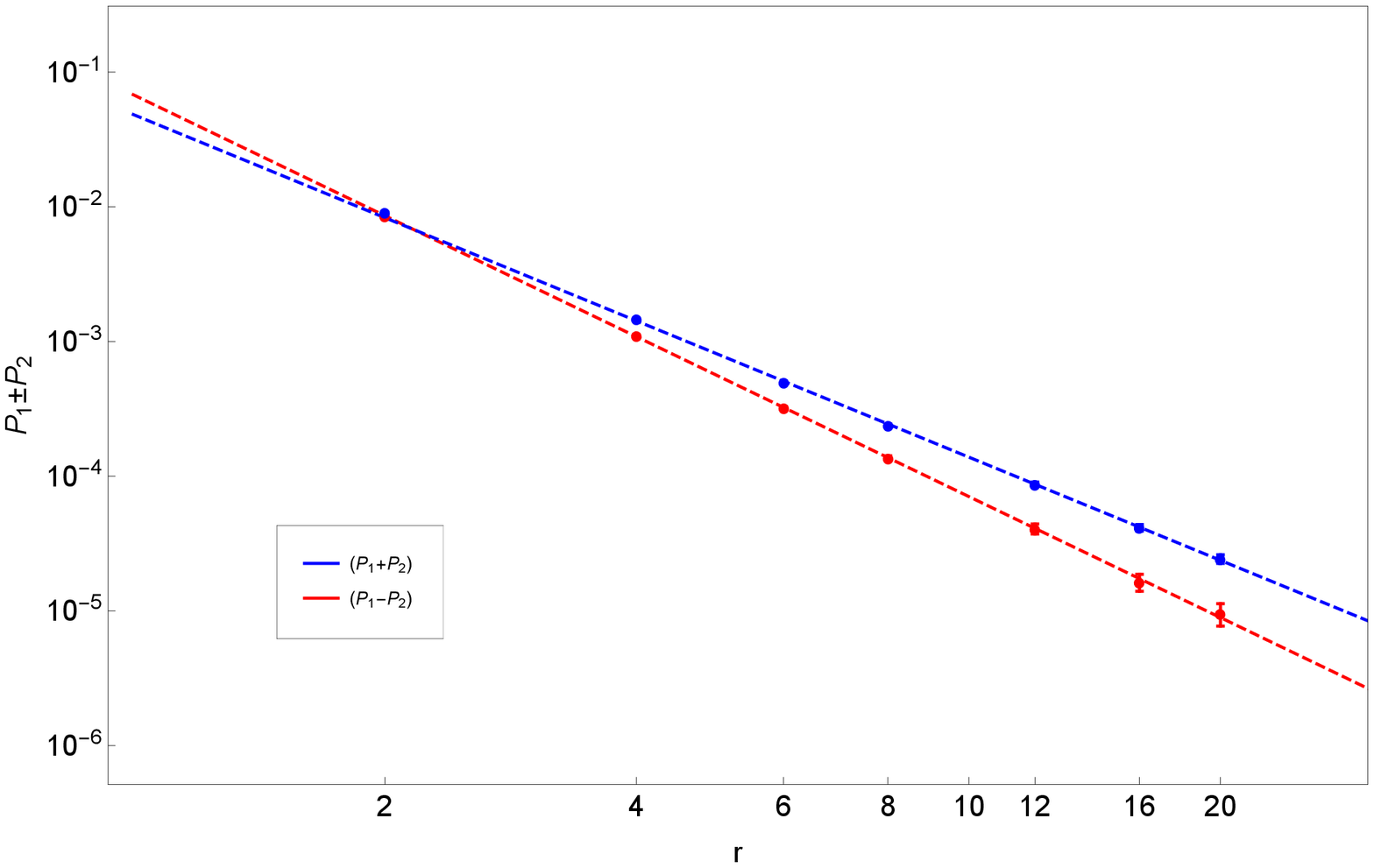}
\caption{Linear combination $P = P_1 \pm P_2$, where $P_1=\protect\Ps\protect$ and $P_2=\protect\Pc\protect$, of probabilities for the propagation of $N=2$ clusters in percolation ($Q=1$, or $x=2$) obtained in Monte Carlo simulations. The quantity $r$ is the distance between the two operators in lattice spacing. The red plot corresponds to \eqref{2points[2]} and the blue plot to \eqref{2points[1,1]}. The critical exponents correspond to slopes $2 \Delta_{0,2}=\frac{5}{2} = 2.5$ and $2 \Delta_{1,2}=\frac{23}{8} = 2.875$. Numerically we find $2\Delta_{0,2}=2.54(8)$ and $2\Delta_{1,2}=2.9(1)$.}
\label{montecarloexp}
\end{figure}

We first investigate the case of percolation, corresponding to $Q=1$. Using Monte Carlo simulations, we have studied the scaling behavior of the $N=2$ two-points function of the tensors $t^{[Q-2,2]}$ and $t^{[Q-2,1,1]}$. The scaling dimensions of the associated fields are predicted by \eqref{Delta_pN} to be $\Delta_{s}=2h_{0,2}=5/4$ and $\Delta_{a}=h_{1/2,2}+h_{-1/2,2}=23/16$ for the symmetric and anti-symmetric tensor, respectively. The averaging was done over $10^8$ configurations. The results, shown in Figure~\ref{montecarloexp}, are in good agreement with the exact values. 

In order to measure the exponents for tensors of higher order, the transfer matrix formalism is helpful. By exact diagonalisation of transfer matrices on a cylinder, we are able to extract the conformal data from finite-size scaling corrections to the free energies \cite{C84,BN82}. We use the Fortuin-Kasteleyn cluster representation. The state for a row of $L$ spins within a time slice is encoded by specifying the way in which the spins are connected through the (parts of) FK clusters constructed at previous times \cite{BN82}. The transfer matrix adds a row to the system. In order to take into account the non-local weight $Q$ for each cluster, the transfer matrix multiplies by the Boltzmann weight $Q$ whenever it acts on the last (in the time direction) spin of a cluster. 

To compute interesting quantities, we place ourselves in sectors where $N \ge 1$ clusters are marked. The marked clusters are constrained to propagate at every step of the action of the transfer matrix, so they cannot be left behind by the time evolution. This means that $N$ distinct clusters propagate from one end to the other of the cylinder. The case $N=1$ corresponds to the magnetisation, and the finite-size scaling of the free energy in this sector provides an estimate of the magnetisation exponent. In general, we extract estimates of critical exponents from the finite-size scaling of the largest eigenvalues in each sector, following \cite{C84}. We have only considered the critical exponents of primaries (i.e., $N=n$) where each marked spin is connected to a propagating cluster.

For a state with $N$ marked clusters, the Young symmetriser acts on the labeling of the clusters. This provides a state in the $N$-cluster sector of the transfer matrix with a well-defined $\Sc_Q$ symmetry. Since the Young symmetriser is idempotent and commutes with the transfer matrix, we can avoid numerical instabilities by letting it act every time a row has been added by the transfer matrix.

For instance, consider a state with $N=2$ marked clusters, $A$ and $B$, that we denote symbolically as $v=\left| X,X,A,B,X,X \right\rangle$. The vector $v$ corresponds to a state of width $L=6$ sites, where the third and fourth spins belong respectively to the clusters marked $A$ and $B$, and the others are not specified in our notation ($X$). They can be initialised, for instance, so that each $X$ corresponds to a different unmarked cluster. The action of the anti-symmetriser \eqref{tensor2antisym} gives the vector $v=\left| X,X,A,B,X,X \right>- \left| X,X,B,A,X,X \right>$ that generates a sector of the transfer matrix. 

\begin{figure}[h!]
\centering
\includegraphics[width=14cm]{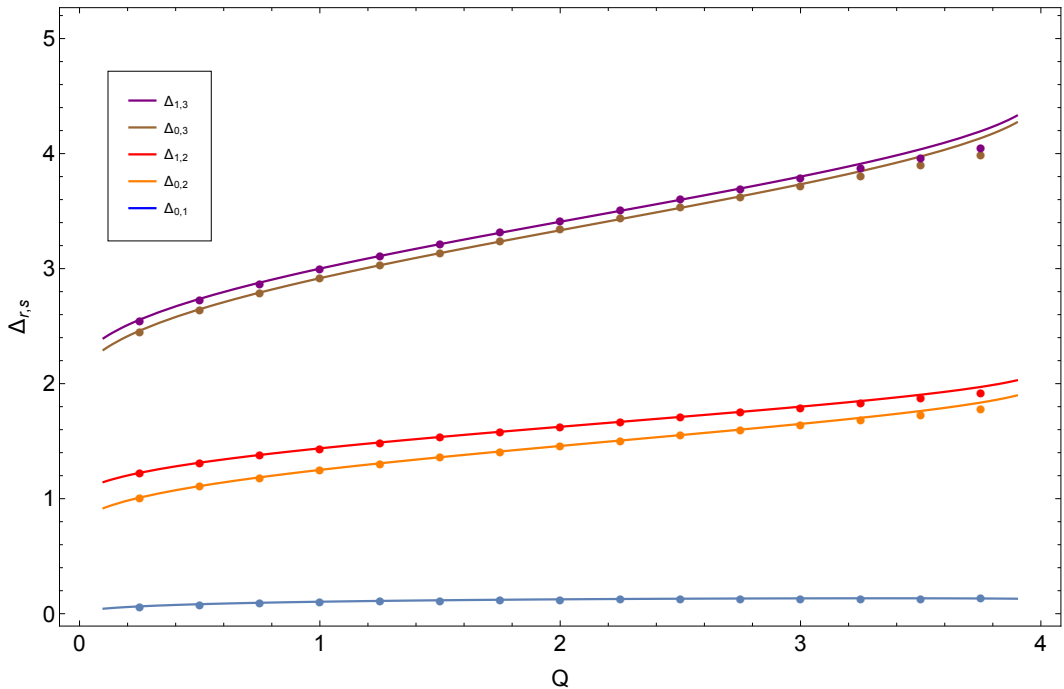}
\caption{Critical exponents for the propagation of two anti-symmetric clusters, as a function of $Q$. We computed all conformal dimensions corresponding to the propagation of $N=1,2,3$ FK-clusters. The lines are the theoretical values $\Delta_{p,N}=h_{p/N,N}+h_{-p/N,N}$ for $N>1$ and 
$\Delta_{0,1}=2h_{1/2,0}$ for the magnetisation operator. The points are the data extrapolated from a numerical diagonalisation of the transfer matrix using finite-size scaling on a cylinder. We observe a very good agreement, as long as we do not take $Q$ too close to $4$. In the latter case the convergence is impeded by logarithmic terms appearing in the finite-size corrections \cite{C86log}.}
\label{transfermatrixexp}
\end{figure}

The largest eigenvalue of each sector is then computed from a standard iterative scheme, and we can extract the scaling dimension of the associated operator. We are able to extract very precise numerical data using this formalism. The numerical extrapolation for the critical exponents, as function of $Q$, is shown in Figure~\ref{transfermatrixexp}.

\subsection{Spin}

It is also possible to verify the predictions for the conformal spin in numerical simulations.

\begin{figure}[!tbp]
  \centering
  \begin{minipage}[b]{0.4\textwidth}
    \psset{unit=0.7cm}
\begin{pspicture}[shift = 0.5](0,0)(10,8)

\multirput(1,0)(1,0){10}{\psline[linewidth = .7pt,linecolor = black](0,0)(0,6)}
\multirput(0,1)(0,1){5}{\psline[linewidth = .7pt,linecolor = black](0,0)(11,0)}

\pscircle[fillstyle=solid, fillcolor=red](3.,3.){0.1}
\pscircle[fillstyle=solid, fillcolor=red](5.,4.0){0.1}

\rput(4,5){\psline[linewidth = .7pt,linecolor = black](-0.1,-0.1)(0.1,0.1)\psline[linewidth = .7pt,linecolor = black](0.1,-.1)(-0.1,.1)}
\rput(1,4){\psline[linewidth = .7pt,linecolor = black](-0.1,-0.1)(0.1,0.1)\psline[linewidth = .7pt,linecolor = black](0.1,-.1)(-0.1,.1)}
\rput(1,2){\psline[linewidth = .7pt,linecolor = black](-0.1,-0.1)(0.1,0.1)\psline[linewidth = .7pt,linecolor = black](0.1,-.1)(-0.1,.1)}
\rput(2,1){\psline[linewidth = .7pt,linecolor = black](-0.1,-0.1)(0.1,0.1)\psline[linewidth = .7pt,linecolor = black](0.1,-.1)(-0.1,.1)}
\rput(4,1){\psline[linewidth = .7pt,linecolor = black](-0.1,-0.1)(0.1,0.1)\psline[linewidth = .7pt,linecolor = black](0.1,-.1)(-0.1,.1)}
\rput(5,2){\psline[linewidth = .7pt,linecolor = black](-0.1,-0.1)(0.1,0.1)\psline[linewidth = .7pt,linecolor = black](0.1,-.1)(-0.1,.1)}
\rput(2,5){\psline[linewidth = .7pt,linecolor = black](-0.1,-0.1)(0.1,0.1)\psline[linewidth = .7pt,linecolor = black](0.1,-.1)(-0.1,.1)}

\pscircle[fillstyle=solid, fillcolor=blue](8.,3.){0.1}
\pscircle[fillstyle=solid, fillcolor=blue](9.,3.0){0.1}

\psline[linewidth = .5pt,linecolor = black,linestyle=dashed](3,3)(5,4)
\psarc(3,3){0.5}{0}{28}
\rput(3.8,3.2){\footnotesize$\theta$}

\end{pspicture}
   
  \end{minipage}
\hspace{1.5cm}
  \begin{minipage}[b]{0.4\textwidth}

\psset{unit=0.7cm}
\begin{pspicture}[shift = 0](0,0)(10,8)

\multirput(1,0)(1,0){10}{\psline[linewidth = .7pt,linecolor = black](0,0)(0,7)}
\multirput(0,1)(0,1){6}{\psline[linewidth = .7pt,linecolor = black](0,0)(11,0)}

\pscircle[fillstyle=solid, fillcolor=red](2.,2.){0.1}
\pscircle[fillstyle=solid, fillcolor=red](3.,2.0){0.1}
\psline[linewidth = .7pt,linecolor = black](2.4,1.9)(2.6,2.1)
\psline[linewidth = .7pt,linecolor = black](2.4,2.1)(2.6,1.9)

\pscircle[fillstyle=solid, fillcolor=blue](8.,2.){0.1}
\pscircle[fillstyle=solid, fillcolor=blue](9.,2.0){0.1}
\psline[linewidth = .7pt,linecolor = black](8.4,1.9)(8.6,2.1)
\psline[linewidth = .7pt,linecolor = black](8.4,2.1)(8.6,1.9)

\pscircle[fillstyle=solid, fillcolor=blue](7.,5.){0.1}
\pscircle[fillstyle=solid, fillcolor=blue](8.,5.){0.1}
\psline[linewidth = .7pt,linecolor = black](7.4,4.9)(7.6,5.1)
\psline[linewidth = .7pt,linecolor = black](7.4,5.1)(7.6,4.9)

\psline[linewidth = .5pt,linecolor = black,linestyle=dashed](2.5,2)(7.5,5)
\psarc(2.5,2){0.8}{0}{31}
\rput(3.5,2.3){\footnotesize$\theta$}

\end{pspicture}

  \end{minipage}
  \caption{Rotations performed on the lattice to measure the spin of an operator acting on $N=2$ sites. {\em Left panel}\/: One operator acts on the two red sites. We measure the correlation function with another operator (corresponding to the two blue sites) with Monte Carlo simulations. We can move the right red dot around the other one to measure the spin of the operator through the correlation function but this is very restricted by the lattice discretisation (the usable positions are marked with a cross). A first solution to this issue is to increase the distance between the two red sites (this is allowed if the two red points are close enough compared to the distance with the blue sites). {\em Right panel}\/: A simpler solution is to keep the relative orientation of each pair of sites fixed and move globally the blue sites around the red ones. This rotation defines the angle $\theta$. Those correlations are easier to measure, since we are less affected by the lattice discretisation.}
\label{globalrotation}
\end{figure}
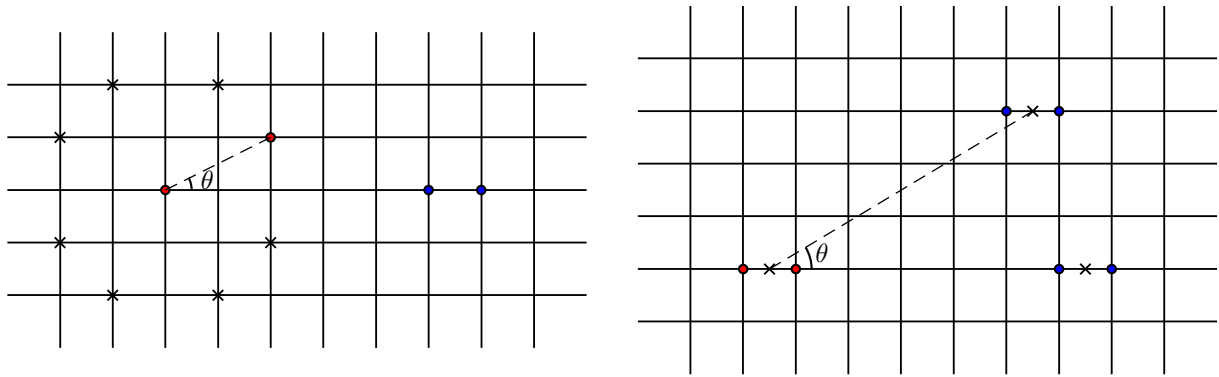

Let us consider, for instance, the two-point function corresponding to the tensor \eqref{tensor2antisym}. From the previous section, we expect that $t^{[Q-2,1,1],[1,1]}_{a_1,a_2}$ corresponds, in $d=2$ CFT, to a field of spin $1$. It acts on two sites in the same neighbourhood, whose relative orientation defines a unit vector $u^\mu$, which corresponds to the direction of the vector going from the first to the second site. The general correlation function of a vector field $\Op_\mu$ is predicted by conformal invariance to be \cite{OP94,Rychkov11}
\be \left<\Op_\mu(x)\Op_\nu(y)\right>=\frac{\dt_{\mu,\nu}-2\frac{(x-y)_\mu (x-y)\nu}{(x-y)^2}}{|x-y|^{2\Delta}} \,. \ee
We define also $\Op_u=u^\mu\Op_\mu$. Thus, the correlation function evaluated at the points $x=0$ and $y=(r,0)$ reads
\be \left<\Op_{u_1}(x)\Op_{u_2}(y)\right>=-\frac{\cos(\theta_1+\theta_2)}{r^{2\Delta}} \,, \ee
where $u_1=(\cos\theta_1,\sin\theta_1)$ and $u_2=(\cos\theta_2,\sin\theta_2)$.

\begin{figure}[h!]
\centering
\includegraphics[width=14cm]{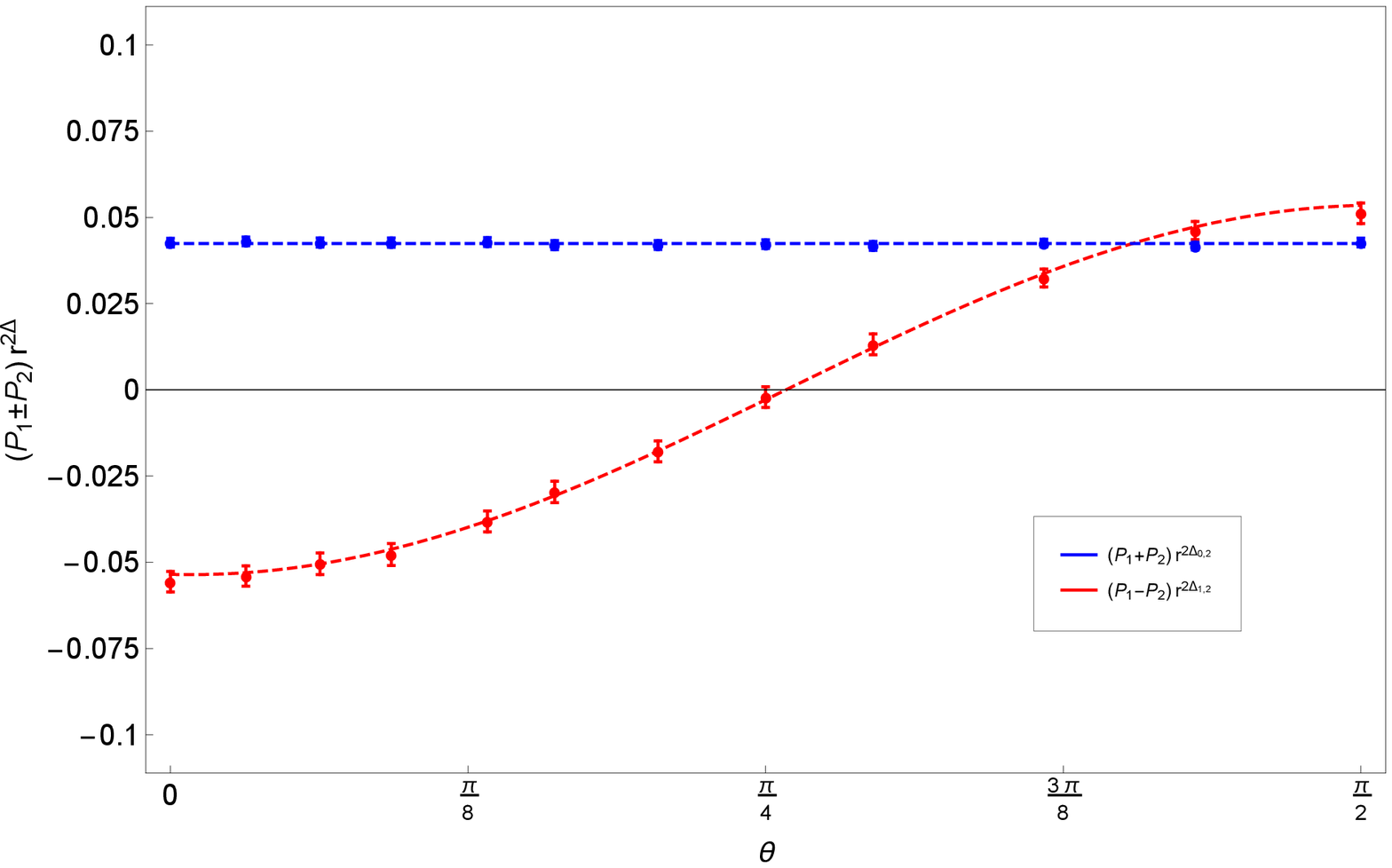}

\caption{Linear combination $P = P_1 \pm P_2$, where $P_1=\protect\Ps\protect$ and $P_2=\protect\Pc\protect$, of renormalised probabilities for the propagation of $N=2$ clusters in percolation ($Q=1$, or $x=2$). A rotations corresponding to the one described on the right panel of Figure \ref{globalrotation} is performed and parametrized by $\theta$. The red plot corresponds to \eqref{2points[2]} and the blue plot to \eqref{2points[1,1]}, whose angular dependence is proportional to $1$ and $\cos 2\theta$, respectively.}
\label{spinfigure}
\end{figure}

It is readily shown that if we keep $u_1$ and $u_2$ fixed, and rotate $x$ around $y$ through an angle $\theta$, the above expression is multiplied by an angular factor $\cos2\theta$. This last property is amenable to numerical verification. Because of the lattice discretisation, it is in fact easier to move one of the points, take the orientation fixed (Figure~\ref{globalrotation}), rather than to apply a local rotation. Figure \ref{spinfigure} shows that we can measure this rotation effect on the lattice using Monte Carlo simulations.

\section{Conclusion}

In this paper, following the work of two of the authors \cite{VJ2014}, we have generalised the classification of operators in the Potts model, by using its underlying $\Sc_Q$ symmetry. The initial inspiration to this seris of works was Cardy's ideas \cite{C99} to use discrete symmetries and a limiting procedure to study LCFTs. Here we obtained a large set of non-scalar operators, completing the scalar ones found in \cite{VJ2014}.

We computed their correlation functions and, using this, we identified observables on the lattice that behave non-trivially under rotations. The analysis in the two-dimensional case revealed that the corresponding critical exponents take known values \cite{ReadSaleur01}. We were able to verify their conformal dimensions using transfer-matrix diagonalisation and Monte Carlo methods. In addition, we discussed the meaning of the conformal spin for lattice observables.

Our techniques are independent of the dimension $d$, and we believe that our classification defines a family of operators corresponding to quasi-primaries for $d>2$. Those operators are supposed to mix into Jordan cells for the dilatation operator in any dimension, for certain (integer) values of $Q$ that can be inferred from our conjecture \eqref{conjecture}.

\ack
We thank Hubert Saleur for inspiring discussions and collaboration on related subjects.
 
This research of was supported in part by the Institut
Universitaire de France (JLJ), and the European Research Council, through the 
advanced grant NuQFT (RC and JLJ).
We also acknowledge support from the Department of Energy through the LDRD program of LBNL (RV).

\appendix

\section{Operators acting on $3$ spins}
\label{sec:app3spins}

\subsection{Derivations of all tensors}

In this section we consider the case of operators acting on $N=3$ spins within all possible representations of $\Sc_N$. We call $M_Q^{(3,3)}$ the space of operators acting on $3$ different spins. The next subsections define and discuss the spaces $M_Q^{(N,k)}$ of operators that act on more spins. For $N=3$ we have to consider $3$ different diagrams: $[3]$, $[2,1]$ and $[1,1,1]$. 

\subsubsection{Case of $\lambda_N = [3]$.}

In the case of the fully symmetric representation $[3]$, there are $4$ corresponding tensors with respectively $\Sc_Q$ representations $\lambda_Q = [Q]$, $[Q-1,1]$, $[Q-2,2]$ and $[Q-3,3]$. We do not compute them here, since they have already been explicitly written in \cite{VJ2014}. In the space $ M_Q^{(3,3)}$ they contribute to $4$ different subspaces in the decomposition \eqref{MQN3} below. 

\subsubsection{Case of $\lambda_N = [2,1]$.}

We thus start with the $\Sc_N$ representation $[2,1]$. To this end, 
we consider the following normal tableau 
\be
\ytableausetup{boxsize=2em,centertableaux}
\begin{ytableau}
 1 & 2  \\
 3
\end{ytableau}
\label{firstmixed}
\ee

The second normal tableau (with $2$ and $3$ permuted) gives another tensor that is equivalent (but independent). We now compute the operators in all the different possible representations of $\Sc_Q$ that are compatible. We start with the diagram $\la_Q=[Q-1,1]$ since $[Q]$ does not satisfy the consistency rule. We take the following tableau
\be
\ytableausetup{boxsize=2em,centertableaux}
\begin{ytableau}
 i_1 & i_2 & \ldots & \, i_{Q-1} \, \\
 a
\end{ytableau}
\ee
and define $t_a^{[Q-1,1],[2,1]}$ using \eqref{definitionTensor}:
\bea t_a^{[Q-1,1],[2,1]} &=&e_{\la_Q}^{(a)}\sn{e}_{\la_N}^{(a)}\Op_{a,i_{Q-2},i_{Q-1}}\nonumber\\
&=&e_{\la_Q}^{(a)}\left(\Op_{a,i_{Q-2},i_{Q-1}}+\Op_{i_{Q-2},a,i_{Q-1}}-\Op_{i_{Q-1},a,i_{Q-2}}-\Op_{i_{Q-1},i_{Q-2},a}\right)\nonumber\\
&=&\left(\sum_{h\in \tilde{h}_{\la_Q}^{(a)}}h\right)\Big(\Op_{a,i_{Q-2},i_{Q-1}}+\Op_{i_{Q-2},a,i_{Q-1}}-\Op_{i_{Q-1},a,i_{Q-2}}-\Op_{i_{Q-1},i_{Q-2},a}\nonumber\\&&-\Op_{i_1,i_{Q-2},i_{Q-1}}-\Op_{i_{Q-2},i_1,i_{Q-1}}+\Op_{i_{Q-1},i_1,i_{Q-2}}+\Op_{i_{Q-1},i_{Q-2},i_1}\Big)
\eea
where again $\tilde{h}_{\la_Q^{(a)}}$ is the set of permutations acting only on the first row of $\la_Q$. After the symmetrisation a lot of terms are cancelled and we obtain (after a normalisation) :
\bea t_a^{[Q-1,1],[2,1]} = \sum_{\substack{i,j=1\\i\neq j\\i,j\neq a}}^Q\left(\Op_{a,i,j}-\Op_{i,j,a}\right) .\eea

If we choose the other possible Young tableau for $\la_N=[2,1]$, which is 
\be
\ytableausetup{boxsize=2em,centertableaux}
\begin{ytableau}
 1 & 3  \\
 2
\end{ytableau}
\ee
we obtain the following tensor 
\bea t_a^{[Q-1,1],[2,1]'} = \sum_{\substack{i,j=1\\i\neq j\\i,j\neq a}}^Q\left(\Op_{a,i,j}-\Op_{i,a,j}\right) .\eea
Those two possible tensors are independent and each contribute to a representation $[Q-1,1]$ in the decomposition of $M_Q^{(N,3)}$. In the rest of this section we only consider \eqref{firstmixed}; the operators for the second normal tableau are computed similarly.

The next diagram to consider, and which is consistent with the mixed $[2,1]$ representation, is $\la_Q=[Q-2,2]$. We follow our general construction and define the tableau 
\be
\ytableausetup{boxsize=2em,centertableaux}
\begin{ytableau}
 i_1 & i_2 & \ldots & \, i_{Q-2} \, \\
 a_1 & a_2
\end{ytableau}
\ee
and the tensor $t_{a_1,a_2}^{[Q-2,2],[2,1]}$ is:

\bea t_{a_1,a_2}^{[Q-2,2],\la_N} &=&e_{\la_Q}^{(a)}\sn{e}_{\la_N}^{(a)}\Op_{a_1,a_2,i_{Q-2}}\nonumber\\
&=&e_{\la_Q}^{(a)}\Big(\Op_{a_1,a_2,i_{Q-2}}+\Op_{a_2,a_1,i_{Q-2}}-\Op_{i_{Q-2},a_1,a_2}-\Op_{i_{Q-2},a_2,a_1}\Big)
\eea
A direct computation gives 
\bea t_{a_1,a_2}^{[Q-2,2],\la_N} &=&(Q-3)! \sum_{\substack{i=1\\i\neq a_1,a_2}}^Q\left(\Op_{a_1,a_2,i}+\Op_{a_2,a_1,i}-\Op_{i,a_1,a_2}-\Op_{i,a_2,a_1}\right)\nonumber\\
&+&(Q-4)!\sum_{\substack{i,j=1\\i\neq j\\i,j\neq a_1,a_2}}^Q\left(\Op_{i,j,a_2}+\Op_{i,j,a_1}-\Op_{a_1,i,j}-\Op_{a_2,i,j}\right) \,,
\eea
which can be written in the following form (after normalisation):
\bea t_{a_1,a_2}^{[Q-2,2],\la_N} &=& \sum_{\substack{i=1\\ i\neq a_1,a_2}}^Q\left(\Op_{a_1,a_2,i}+\Op_{a_2,a_1,i}-\Op_{i,a_1,a_2}-\Op_{i,a_2,a_1}\right)\nonumber\\&&-\tfrac{1}{Q-2}\left(t_{a_1}^{[Q-1,1],[2,1]}+t_{a_2}^{[Q-1,1],[2,1]}\right). \eea
The tensor is indeed invariant under the permutation of its two indices $a_1$ and $a_2$.

The case of the representation with Young diagram $\la_Q=[Q-2,1,1]$ is very similar. We start from the Young tableau
\be
\ytableausetup{boxsize=2em,centertableaux}
\begin{ytableau}
 i_1 & i_2 & \ldots & \, i_{Q-2} \, \\
 a_1 \\
 a_2
\end{ytableau}
\ee
and define  $t_{a_1,a_2}^{[Q-2,1,1],[2,1]}$ as previously. We find that it takes the following form in terms of  $t_{a}^{[Q-1,1],[2,1]}$:
\bea t_{a_1,a_2}^{[Q-2,1,1],\la_N} &=&\sum_{\substack{i=1\\ i\neq a_1,a_2}}^Q\left(\Op_{a_1,a_2,i}-\Op_{a_2,a_1,i}+2\Op_{a_1,i,a_2}-2\Op_{a_2,i,a_1}+\Op_{i,a_1,a_2}-\Op_{i,a_2,a_1}\right)\nonumber\\&&-\tfrac{3}{Q}\left(t_{a_1}^{[Q-1,1]}-t_{a_2}^{[Q-1,1]}\right).\eea
The tensor obtained is obviously anti-symmetric under the permutation of $a_1$ and $a_2$.

Last, but not least, we define the primal operator corresponding to $[Q-3,2,1]$. The computation is tedious but straightforward, and we find
\bea t_{a_1,a_2,a_3}^{[Q-3,2,1]} &=& \Op_{a_1,a_2,a_3}+\Op_{a_2,a_1,a_3}-\Op_{a_3,a_1,a_2}-\Op_{a_3,a_2,a_1}\nonumber\\
&&-\tfrac{1}{2(Q-1)}\left(2 t_{a_1,a_2}^{[Q-2,2],\la_N}- t_{a_1,a_3}^{[Q-2,2],\la_N}- t_{a_2,a_3}^{[Q-2,2],\la_N}\right)\nonumber\\
&&-\tfrac{1}{2(Q-3)}\left(t_{a_1,a_3}^{[Q-2,1,1],\la_N}+t_{a_2,a_3}^{[Q-2,1,1],\la_N}\right)\nonumber\\
&&-\tfrac{1}{Q(Q-2)}\left(t_{a_1}^{[Q-1,1],\la_N}+t_{a_2}^{[Q-1,1],\la_N}-2t_{a_3}^{[Q-1,1],\la_N}\right)
.\eea
We emphasise again the fact that another tensor exists with this symmetry (corresponding to the other normal Young tableau of $\Sc_N$) and explains the factor $2$ in the decomposition \eqref{MQN3}.

\subsubsection{Case of $\lambda_N = [1,1,1]$.}

In order to be complete for $N=3$ spins, we discuss now the case of tensors with $\Sc_N$ symmetry $\la_N=[1,1,1]$. The first adequate $\Sc_Q$ representation that we have to consider is $\la_Q=[Q-2,1,1]$.
We consider the tableau
\be
\ytableausetup{boxsize=2em,centertableaux}
\begin{ytableau}
 i_1 & i_2 & \ldots & \, i_{Q-2} \, \\
 a_1 \\
 a_2
\end{ytableau}
\ee
and compute $t_{a_1,a_2}^{[Q-2,1,1],[1,1,1]}$ such that:
\bea 
t_{a_1,a_2}^{[Q-2,1,1],[1,1,1]}&=&e_{\la_Q}^{(a)}\Big(\Op_{a_1,a_2,i_{Q-2}}+\Op_{a_2,i_{Q-2},a_1}+\Op_{i_{Q-2},a_1,a_1}-\Op_{a_2,a_1,i_{Q-2}}\nonumber\\&&-\Op_{a_1,i_{Q-2},a_2}-\Op_{i_{Q-2},a_2,a_1}\Big)\nonumber\\
&=&\sum_{\substack{i=1\\ i\neq a_1,a_2}}^Q\Big(\Op_{a_1,a_2,i}-\Op_{a_2,a_1,i}+\Op_{i,a_1,a_2}-\Op_{i,a_2,a_1}\nonumber\\&&+\Op_{a_2,i,a_1}-\Op_{a_1,i,a_2}\Big) \,,
\eea
where we normalised the tensor so as to remove the factor $(Q-3)!$.

Finally, the last tensor that we have to consider is the primal with $\Sc_Q$ symmetry $\la_Q=[Q-3,1,1,1]$. A computation shows that
\bea t_{a_1,a_2,a_3}^{[Q-3,1,1,1]}&=&\Op_{a_1,a_2,a_3}+\Op_{a_2,a_3,a_1}+\Op_{a_3,a_1,a_2}-\Op_{a_1,a_3,a_2}-\Op_{a_3,a_2,a_1}-\Op_{a_2,a_1,a_3}\nonumber\\
&-&\tfrac{1}{Q}\left( t_{a_1,a_2}^{[Q-2,1,1],[1,1,1]}+ t_{a_2,a_3}^{[Q-2,1,1],[1,1,1]}+ t_{a_3,a_1}^{[Q-2,1,1],[1,1,1]}\right).\eea
\subsection{Space of tensors acting on $3$ spins}
We sum up here the results of last section concerning the space of operators acting on $N=3$ different spins. We denote this space by $M_Q^{(3,3)}$ (see \ref{sec:general-dimension} for more general definitions). It corresponds to $\left\{\Op_{i,j,k}\, : \, i,j,k=1,\ldots,Q \mbox{ and } i\neq j \neq k \right\}$. The dimension of this space is obviously $Q(Q-1)(Q-2)$.

We can decompose this space in terms of the different tensors that we have defined. First let us consider the case of tensors with $\Sc_N$ symmetry corresponding to $\lambda_N = [3]$. There are four different tensors respectively corresponding to the $\Sc_Q$ diagrams $[Q-3,3]$, $[Q-2,2]$, $[Q-1,1]$ and $[Q]$. Next, for the tensors with symmetry $\lambda_N = [2,1]$ there are $4$ compatible $\Sc_Q$ symmetries: $[Q-3,2,1]$, $[Q-2,2]$, $[Q-2,1,1]$ and $[Q-1,1]$. Since there is $2$ different ways to define the tableau of shape $[2,1]$ we can define $2$ independent tensors from each $\Sc_Q$ tableaux. The last $\Sc_n$ tableau, $\lambda_N = [1,1,1]$, is compatible with only $2$ Young diagrams $[Q-3,1,1,1]$ and $[Q-2,1,1]$.

In the end, we obtain the following decomposition 
\bea M_Q^{(3,3)} &=& \left([Q]\oplus[Q-1,1]\oplus[Q-2,2]\oplus[Q-3,3]\right)\nonumber\\&&\oplus2\left([Q-1,1]\oplus[Q-2,2]\oplus[Q-2,1,1]\oplus[Q-3,2,1]\right)\nonumber\\&&\oplus\left([Q-2,1,1]\oplus[Q-3,1,1,1]\right)
\eea
We can check that the result is consistent with the dimensions given by the hook formula (\ref{hook-formula})
\bea
{\rm dim} \, M_Q^{(3,3)} &=& \left(1+(Q-1)+\tfrac{Q(Q-3)}{2}+\tfrac{Q(Q-1)(Q-5)}{6}\right)\nonumber\\
 &&+2\left((Q-1)+\tfrac{Q(Q-3)}{2}+\tfrac{(Q-1)(Q-2)}{2}+\tfrac{Q(Q-2)(Q-4)}{3}\right)\nonumber\\
 &&+\left(\tfrac{(Q-1)(Q-2)}{2}+\tfrac{(Q-1)(Q-2)(Q-3)}{6}\right)\nonumber\\
 &=&Q(Q-1)(Q-2).
\eea
\subsection{Correlation functions}
We now give the explicit form of the two-point functions for the $N=3$ operators described previously. We use the method detailed in section \ref{correlationfunctions} to write the correlation functions in terms of the probabilities describing the propagation of FK-clusters. We start with $t_{a}^{[Q-1,1],[2,1]}$; its two-point function reads
\bea\hspace{-2cm}
\left<t_a^{[Q-1,1],[2,1]}t_b^{[Q-1,1],[2,1]}\right>&=\tfrac{Q-2}{Q^2}\left(\dt_{a,b}-\tfrac{1}{Q}\right)\bigg(\tfrac{Q-2}{Q}\bigg(\resizebox{0.3\hsize}{!}{$\Pannann+\Pnnanna-\Pannnna-\Pnnaann$}\bigg)\nonumber\\
&\hspace{-1cm}+\tfrac{Q-2}{Q}\bigg(\resizebox{0.6\hsize}{!}{$2\Panbanb+\Pabnnab+\Pnababn+\Pnabanb+\Panbnab+\Panbabn+\Pabnanb$}\nonumber\\
&\hspace{-1cm}-\resizebox{0.6\hsize}{!}{$2\Panbbna-\Pnabnba-\Pabnban-\Pabnbna-\Panbnba-\Panbban-\Pnabbna$}\bigg)\nonumber\\
&\hspace{-1cm}\tfrac{Q-1}{Q}\bigg(\resizebox{0.65\hsize}{!}{$\Pabnabn+\Pabnban+\Pnabnab+\Pnabnba-\Pnabban-\Pnababn-\Pabnnba-\Pabnnab$}\bigg)\nonumber\\
&\hspace{-1cm}+\resizebox{0.55\hsize}{!}{$2\Pabc+\Pacb+\Pbac-2\Pcba-\Pbca-\Pcab$}.
\eea
As expected there are two different parts in this correlation. The symmetric group fixes the pre-factor that involves Kronecker symbols of tensorial indices. The most interesting part is the linear combination of probabilities that a certain number of clusters propagate. In the following we omit the pre-factor. The two-point function of the secondary operator $t_{a,b}^{[Q-2,2],[2,1]}$ is
\bea\hspace{-2cm}
\left<t_{a_1,a_2}^{[Q-2,2],[2,1]}t_{b_1,b_2}^{[Q-2,2],[2,1]}\right>\propto&\bigg(\tfrac{Q-1}{Q}\bigg(\resizebox{0.6\hsize}{!}{$\Pabnabn+\Pabnban+\Pnabnab+\Pnabnba-\Pabnnab-\Pabnnba-\Pnababn-\Pnabban$}\bigg)\nonumber\\
&+\resizebox{0.5\hsize}{!}{$2\Pabc+\Pacb+\Pbac-2\Pcba-\Pbca-\Pcab$}\bigg).
\eea
We continue with the operator $t^{[Q-2,1,1],[2,1]}$:
\bea\hspace{-2cm}
\left<t_{a_1,a_2}^{[Q-2,1,1],[2,1]}t_{b_1,b_2}^{[Q-2,1,1],[2,1]}\right>\propto&\bigg(\resizebox{0.5\hsize}{!}{$6\bigg(2\,\Pabc+\Pacb+\Pbac-2\,\Pcba-\Pbca-\Pcab\bigg)$}\nonumber\\
&\hspace{-1cm}\resizebox{0.7\hsize}{!}{$\hspace{-1cm}+\tfrac{Q-3}{Q}\bigg(4\,\Panbanb+2\,\Pabnanb+2\,\Panbabn+2\,\Panbnab+2\,\Pnabanb+\Pabnabn+\Pabnnab+\Pnababn+\Pnabnab$}\nonumber\\
&\hspace{-1cm}\resizebox{0.7\hsize}{!}{$\hspace{-1cm}-4\,\Panbban-2\,\Pabnbna-2\,\Pabnban-2\,\Panbnba-2\,\Pnabbna-\Pabnban-\Pabnnba-\Pnabban-\Pnabnba$}\bigg)\bigg) \,. \nonumber\\.
\eea
The last operator with symmetry $[2,1]$ on the $3$ spins has a two-point function involving only probabilities where $3$ clusters propagate (because it is primal):
\bea\hspace{-2cm}
\left<t_{a_1,a_2,a_3}^{[Q-3,2,1],[2,1]}t_{b_1,b_2,b_3}^{[Q-3,2,1],[2,1]}\right>\propto\bigg(\resizebox{0.55\hsize}{!}{$2\Pabc+\Pacb+\Pbac-2\Pcba-\Pbca-\Pcab$}\bigg) \,. \nonumber \\
\eea

And finally we give the expression in the case where the $3$ spins are antisymmetric $[1,1,1]$. The secondary operator $t^{[Q-2,1,1],[1,1,1]}$ has the following two-point function:
\bea\hspace{-2cm}
\left<t_{a_1,a_2}^{[Q-2,1,1],[1,1,1]}t_{b_1,b_2}^{[Q-2,1,1],[1,1,1]}\right>\propto&\bigg(\resizebox{0.55\hsize}{!}{$3\,\Pabc+3\,\Pbca+3\,\Pcab-3\,\Pcba-3\,\Pbac-3\,\Pacb$}\nonumber\\
&\hspace{-1cm}\resizebox{0.7\hsize}{!}{$\hspace{-1cm}+\Pabnabn+\Pnababn+\Panbban+\Pabnnab+\Pabnbna+\Panbanb+\Panbnba+\Pnabbna+\Pnabnab$}\nonumber\\
&\hspace{-1cm}\resizebox{0.7\hsize}{!}{$\hspace{-1cm}-\Pabnban-\Pnabban-\Panbabn-\Pabnnba-\Pabnanb-\Pnabanb-\Panbnab-\Pnabnba-\Panbbna$}\nonumber\bigg)\\&.
\eea

The primal $t^{[Q-3,1,1,1],[1,1,1]}$ inserts $3$ anti-symmetric clusters:
\bea\hspace{-2cm}
\left<t_{a_1,a_2,a_3}^{[Q-3,1,1,1],[1,1,1]}t_{b_1,b_2,b_3}^{[Q-3,1,1,1],[1,1,1]}\right>\propto\bigg(\resizebox{0.5\hsize}{!}{$\Pabc+\Pbca+\Pcab-\Pcba-\Pbac-\Pacb$}\bigg) \,. \nonumber \\
\eea
\subsection{General results for the dimensions}
\label{sec:general-dimension}
Let us denote by $L_Q^{(N)}$ the space of operators acting on $N$ spins taking any values. Similarly, we denote by $M_Q^{(N,k)}$ the space of operators acting on $N$ spins
constrained to take precisely $k$ distinct values. We have seen that
\bea
 M_Q^{(N,1)} &=& [Q] \oplus [Q-1,1] \label{MQN1} \\
 M_Q^{(N,2)} &=& [Q] \oplus 2 [Q-1,1] \oplus [Q-2,2] \oplus [Q-2,1,1] \\
 M_Q^{(N,3)} &=& [Q] \oplus 3 [Q-1,1] \oplus 3 [Q-2,2] \oplus 3 [Q-2,1,1] \nonumber \\ 
 & & \oplus [Q-3,3] \oplus 2 [Q-3,2,1] \oplus [Q-3,1,1,1] \,. \label{MQN3}
\eea
There are two different sources for the multiplicities in front of each $\Sc_Q$ representation: the number of compatible $\Sc_k$ representations, and the dimension of the
corresponding $\Sc_k$ Young tableau. An $\Sc_k$ representation is compatible if its Young diagram contains the corresponding $\Sc_Q$ Young diagram stripped of its first row, in such a way
that at most one extra box is added to each column.

The sum of dimensions on the right-hand sides of \eqref{MQN1}--\eqref{MQN3} is consistent with the general result
\be
 {\rm dim} \, M_Q^{(N,k)} = Q(Q-1)\cdots(Q-k+1) \equiv (Q)_k \,, \label{dimMQNk}
\ee
where $(Q)_k$ is the falling factorial. We have checked that \eqref{dimMQNk} is verified also for $k=4$ by explicit construction of the decomposition
of $M_Q^{(N,4)}$.

The relation of $M_Q^{(N,k)}$ with the full space $L_Q^{(N)}$ is that it can be obtained by partitioning the $N$-element set into $k$ non-empty sets (called set blocks).
Within each block the corresponding spins are constrained to take identical values. The number of such set partitions is given by the Stirling number of the second
kind
\be
 S(N,k) = \frac{1}{k!} \sum_{i=0}^k (-1)^i {k \choose i} (k-i)^N \,.
\ee
The total dimension therefore correctly comes out as
\be
 {\rm dim} \, L_Q^{(N)} = \sum_{k=1}^N S(N,k) \cdot {\rm dim}\, M_Q^{(N,k)} = Q^N \,.
\ee

\section*{References}
\bibliographystyle{iopart-num}
\bibliography{CJV}

\providecommand{\newblock}{}
\begin{thebibliography}{10}
\expandafter\ifx\csname url\endcsname\relax
  \def\url#1{{\tt #1}}\fi
\expandafter\ifx\csname urlprefix\endcsname\relax\def\urlprefix{URL }\fi
\providecommand{\eprint}[2][]{\url{#2}}

\bibitem{BPZ}
Belavin A~A, Polyakov A~M and Zamolodchikov A~B 1984 {\em Nucl. Phys. B\/} {\bf
  {\bf 241}} 333

\bibitem{FQS}
Friedan D, Qiu Z and Shenker S 1984 {\em Phys. Rev. Lett.\/} {\bf {\bf 52}}
  1575

\bibitem{CardyMin}
Cardy J 1986 {\em Nucl. Phys. B\/} {\bf {\bf 270}} 186

\bibitem{Saleur87}
Saleur H 1987 {\em J. Phys. A: Math. Gen.\/} {\bf {\bf 20}} 455

\bibitem{RS92}
Rozansky L and Saleur H 1992 {\em Nucl. Phys. B\/} {\bf {\bf 376}} 461

\bibitem{Gurarie93}
Gurarie V 1993 {\em Nucl. Phys. B\/} {\bf {\bf 410}} 535

\bibitem{PRZ06}
Pearce P and Zuber J~R~J~B 2006 {\em J. Stat. Mech.\/}  P11017

\bibitem{RS07}
Read N and Saleur H 2007 {\em Nucl. Phys. B\/} {\bf {\bf 777}} 316--351

\bibitem{C99}
Cardy J 1999 {\em arXiv:cond-mat/9911024\/}

\bibitem{LCFT2}
Cardy J 2013 {\em J. Phys. A: Math. Theor.\/} {\bf {\bf 46}} 494001

\bibitem{VJS2012}
Vasseur R, Jacobsen J~L and Saleur H 2012 {\em J. Stat. Mech.\/}  L07001

\bibitem{JSS05}
Jacobsen J~L, Salas J and Sokal A~D 2005 {\em J. Stat. Phys.\/} {\bf {\bf 119}}
  1153--1281

\bibitem{CJSSS04}
Caracciolo S, Jacobsen J~L, Saleur H, Sokal A~D and Sportiello A 2004 {\em
  Phys. Rev. Lett.\/} {\bf {\bf 93}} 080601

\bibitem{JS05}
Jacobsen J~L and Saleur H 2005 {\em Nucl. Phys. B\/} {\bf {\bf 716}} 439--461

\bibitem{DGS07}
Deng Y, Garoni T~M and Sokal A~D 2007 {\em Phys. Rev. Lett.\/} {\bf {\bf 98}}
  030602

\bibitem{CSS17}
S~Caracciolo A~D~S and Sportiello A 2017 {\em J. Phys. A: Math. Theor.\/} {\bf
  {\bf 50}} 114001

\bibitem{FK}
Fortuin C~M and Kasteleyn P~W 1972 {\em Physica\/} {\bf {\bf 57}} 536

\bibitem{BKW76}
Baxter R~J, Kelland S~B and Wu F~Y 1976 {\em J. Phys. A: Math. Gen.\/} {\bf
  {\bf 9}} 397

\bibitem{TL71}
Temperley H~N~V and Lieb E~H 1971 {\em Proc. Roy. Soc. London A\/} {\bf {\bf
  322}} 251--280

\bibitem{VJ2014}
Vasseur R and Jacobsen J~L 2014 {\em Nucl. Phys. B\/} {\bf {\bf 880}} 435--475

\bibitem{HPV16}
Hogervorst M, Paulos M and Vichi A 2016 {\em arXiv:1605.03959\/}

\bibitem{A76}
Amit D~J 1976 {\em J. Phys. A: Math. Gen.\/} {\bf \bf 9} 1441

\bibitem{Baxter73}
Baxter R~J 1973 {\em J. Phys. C: Solid State Phys.\/} {\bf {\bf 6}} L445

\bibitem{WKT}
Tung W~K 1985 {\em Group theory in physics\/} (World Scientific Publishing
  Company)

\bibitem{S89}
Stembridge J 1989 {\em Pacific J. Math.\/} {\bf {\bf 140}} 353--396

\bibitem{ReadSaleur01}
Read N and Saleur H 2001 {\em Nucl. Phys. B\/} {\bf {\bf 613}} 409

\bibitem{GRSV}
Gainutdinov A~M, Read N, Saleur H and Vasseur R 2015 {\em J. High Energy.
  Phys.\/} {\bf {\bf 5}} 114

\bibitem{C84}
Cardy J 1984 {\em J. Phys. A: Math. Gen.\/} {\bf {\bf 19}}

\bibitem{BN82}
Bl\"ote H~W~J and Nightingale M~P 1982 {\em Physica A\/} {\bf {\bf 112}}
  405--465

\bibitem{C86log}
Cardy J 1986 {\em J. Phys. A: Math. Gen.\/} {\bf {\bf 19}}

\bibitem{OP94}
Osborn H and Petkos A 1994 {\em Ann. Phys.\/} {\bf {\bf 231}} 311--362

\bibitem{Rychkov11}
Costa M~S, Penedones J, Poland D and Rychkov V~S 2011 {\em JHEP\/} {\bf {\bf
  1111}} 071

\end{thebibliography}

\end{document}